\documentclass[aps,pre,superscriptaddress,floatfix,twocolumn,10pt]{revtex4-1}

\usepackage{graphicx}   
\usepackage{amssymb}
\usepackage{amsmath}
\begin{document}

\title{Probing the network structure of health deficits in human aging}
\author{Spencer G. Farrell}	
\email{spencer.farrell@dal.ca}	
\affiliation{Dept. of Physics and Atmospheric Science, Dalhousie University, Halifax, Nova Scotia, Canada B3H 4R2}
\author{Arnold B. Mitnitski}	
\email{arnold.mitnitski@dal.ca}
\affiliation{Department of Medicine, Dalhousie University, Halifax, Nova Scotia, Canada  B3H 2Y9}
\author{Olga Theou}	
\email{olga.theou@dal.ca}
\affiliation{Department of Medicine, Dalhousie University, Halifax, Nova Scotia, Canada  B3H 2Y9}
\author{Kenneth Rockwood}	
\email{kenneth.rockwood@dal.ca}
\affiliation{Department of Medicine, Dalhousie University, Halifax, Nova Scotia, Canada  B3H 2Y9}
\affiliation{Division of Geriatric Medicine, Dalhousie University, Halifax, Nova Scotia, Canada B3H 2E1}
\author{Andrew D. Rutenberg}	
\email{andrew.rutenberg@dal.ca}
\affiliation{Dept. of Physics and Atmospheric Science, Dalhousie University,  Halifax, Nova Scotia, Canada B3H 4R2}

\date{\today}
\begin{abstract} 
We confront a network model of human aging and mortality in which nodes represent health attributes that interact within a scale-free network topology, with observational data that uses both clinical and laboratory (pre-clinical) health deficits as network nodes. We find that individual health attributes exhibit a wide range of mutual information with mortality and that, with a reconstruction of their relative connectivity, higher-ranked nodes are more informative. Surprisingly, we find a broad and overlapping range of mutual information of laboratory measures as compared with clinical measures. We confirm similar behavior between most-connected and least-connected model nodes, controlled by the nearest-neighbor connectivity. Furthermore, in both model and observational data, we find that the least-connected (laboratory) nodes damage earlier than the most-connected (clinical) deficits. A mean-field theory of our network model captures and explains this phenomenon, which results from the connectivity of nodes and of their connected neighbors.  We find that other network topologies, including random, small-world, and assortative scale-free networks, exhibit qualitatively different behavior. Our disassortative scale-free network model behaves consistently with our expanded phenomenology observed in human aging, and so is a useful tool to explore mechanisms of and to develop new predictive measures for human aging and mortality. 
\end{abstract}
\maketitle

\section{Introduction} 
Accumulation of damage causes organismal aging \cite{Kirkwood:2005}.   Even in model organisms, with controlled environment and genotype, there are large individual variations in lifespan and in the phenotypes of aging \cite{Herndon:2002, *Kirkwood:2002}. While  many mechanisms cause specific cellular damage \cite{Lopez-Otin:2013}, no single factor  fully controls the process of aging.  This suggests that the aging process is stochastic and results from a variety of damage mechanisms. 

The variability of individual damage accumulation results in differing trajectories of individual health and in differing individual life-spans, and is a fundamental aspect  of individual aging. A simple method of quantifying this individual damage is the Frailty Index (FI) \cite{Mitnitski:2001, Searle:2008}. The FI is the proportion of age-related health issues (``deficits'') that a person has out of a collection of health attributes. The FI is used as a quantitative tool in understanding the health of individuals as they age.  There have been hundreds of papers using an FI based on self-report or clinical data, both for humans \cite{Rockwood:2017} and for animals \cite{Kane:2017}.  Individuals typically accumulate deficits as they age, and so the FI increases with age across a population. The FI captures the heterogeneity in individual health and is predictive of both mortality and other health outcomes \cite{Kulminski:2007, *Evans:2014, *Mitnitski:2016, *Kojima:2018}. 

In previous work  we developed a stochastic network model of aging with damage accumulation \cite{Taneja:2016, Farrell:2016}.  Each individual is modeled as a network of interacting nodes that represent health attributes. Both the nodes and their connections are idealized and do not specify particular health aspects or mechanisms. Connections (links) between neighboring nodes in the network can be interpreted as influence between separate physiological systems.  In our model, damage facilitates subsequent damage of connected nodes. We do not specify the biological mechanisms that cause damage, only that damage rates depend on the proportion of damaged neighbors. Damage promotes more damage and lack of damage facilitates repair. Rather than model the specific biological mechanisms of aging, we model how damage to components of generic physiological systems can accumulate and propagate throughout an organism --- ending with death. 

Even though our model includes no explicit age-dependence in damage rates or mortality, it  captures Gompertz's law of mortality \cite{Gompertz:1825, *Kirkwood:2015}, the average rate of FI accumulation \cite{Mitnitski:2001, Mitnitski:2013}, and the broadening of FI distributions with age \cite{Rockwood:2004, Gu:2009}. By including a false-negative attribution error (i.e. a finite sensitivity) \cite{Farrell:2016}, we can also explain an empirical maximum of observed FI values --  typically between $0.6 - 0.8$ \cite{Searle:2008, Gu:2009, Rockwood:2004, Mitnitski:2013, Hubbard:2015, Bennett:2013}. This shows that age-dependent ``programming'' of either mortality or damage rates are not necessary to explain these features \cite{Kirkwood:2005}. 

We had chosen the Barab\'asi-Albert (BA) preferential attachment algorithm \cite{Barabasi:1999} to generate our scale-free network, both due to the simplicity of the BA algorithm and due to the numerous examples of these scale-free networks in biological systems \cite{Barabasi:2009}. While we had constrained the scale-free network parameters with the available phenomenology, we did not examine whether other common network structures could also recover the same phenomenology. More specifically, we did not identify which observable behavior sensitively depends on the network structure. 
 
Ideally, we could directly reconstruct the network from available data. However, the direct assessment of node connectivity from observational data is a challenging and generally unsolved problem. Nevertheless, we show here that we can reliably reconstruct the relative connectivity (i.e. the rank-order) of high degree nodes in both model and in large-cohort observational data by measuring mutual dependence between pairs of nodes. This reconstruction allows us to qualitatively confirm the relationship between the connectivity of nodes and how informative they are about mortality \cite{Farrell:2016}. Specifically, we demonstrate that a network with a wide range of node connectivities (such as a scale-free network) is needed to describe the observational data. 

Recently, the FI approach has been extended to laboratory \cite{Howlett:2014} and biomarker data \cite{Mitnitski:2015} and used in clinical \cite{Klausen:2017, *King:2017} and population settings \cite{Blodgett:2017}. Two different FIs have been constructed to measure different types of damage, $F_{\mathrm{clin}}$, with clinically evaluated or self-reported data, and $F_{\mathrm{lab}}$, with lab or biomarker data. Clinical deficits are typically based on disabilities, loss of function, or diagnosis of disease, and they measure clinically observable damage that typically occurs late in life. Lab deficits or biomarkers use the results of lab tests (e.g. blood tests or vital signs) that are binarized using standard reference ranges \cite{McPherson:2016}. Since frailty indices based on laboratory tests measure pre-clinical damage, they are distinct from those based on clinical and/or self-report data \cite{Howlett:2014, Blodgett:2017}.

Even though they measure very different types of damage, both FIs are similarly associated with mortality \cite{Blodgett:2016, Howlett:2014}. Earlier observational studies have found (average) $\langle F_{\mathrm{lab}} \rangle$ larger than $\langle F_{\mathrm{clin}} \rangle$ \cite{Blodgett:2016, Howlett:2014, Mitnitski:2015}. However, a study  of older long-term care patients has found $\langle F_{\mathrm{lab}} \rangle$ less than $\langle F_{\mathrm{clin}} \rangle$ \cite{Rockwood:2015}. While differences between studies could be attributed to classification differences, a large single study  including ages from 20-85 from the National Health and Nutrition Examination Survey (NHANES) \cite{Blodgett:2017} also found that $\langle F_{\mathrm{lab}} \rangle$ was higher than $\langle F_{\mathrm{clin}} \rangle$ at earlier ages, but below at later ages. 

The observed age-dependent relationship (or ``age-structure'') between $F_{\mathrm{lab}}$ and $F_{\mathrm{clin}}$ challenges us to examine whether network properties can determine similar age-structure in model data. We aim to determine what qualitative network features are necessary to explain age-structure. Our working hypothesis is that low-degree nodes should correspond to $F_{\mathrm{lab}}$, just as high-degree nodes correspond to $F_{\mathrm{clin}}$ \cite{Taneja:2016, Farrell:2016}.

Complex networks have structural features beyond the degree distribution. For example, nearest-neighbor degree correlations describe how connections are made between specific nodes of different degree \cite{Barabasi:2016}. Accordingly, we consider networks with three types of degree correlations: assortative, disassortative, and neutral \cite{Barabasi:2016, Newman:2002}.  Networks with assortative correlations tend to connect like-degree nodes, those with disassortative correlations tend to connect unlike-degrees, and those with neutral correlations are random. We probe and understand the internal structure of these networks by examining $F_{\mathrm{high}}$ and $F_{\mathrm{low}}$, i.e. damage to high degree nodes and damage to low degree nodes. 

Since networks have many properties other than degree distribution and nearest-neighbor degree correlations, we have also constructed a mean-field theory that {\em only} has these properties. With it we can better connect specific network properties with qualitatively observed phenomenon, within the context of our network model.

We show how network properties of degree distribution and degree correlations are essential for our model to recover results from observational data. Doing so, we can explain how damage propagates through our network and what makes nodes informative of mortality. This allows us to understand the differences between $F_{\mathrm{low}}$ and $F_{\mathrm{high}}$, or between pre-clinical and clinical damage in observational health data.

\section{Methods}

\subsection{Stochastic model} 
Our model was previously presented \cite{Farrell:2016}. Individuals are represented as a network consisting of $N$ nodes, where each node $i \in \{1, 2, \ldots, N \}$ can take on binary values $d_i = 0,1$ for healthy or damaged, respectively. Connections are undirected and all nodes are undamaged at time $t = 0$. 

A stochastic process transitions between healthy and damaged ($d_i = 0,1$) states. Healthy nodes damage with rate $\Gamma_+ = \Gamma_0 \exp{(f_i \gamma_+)}$ and damaged nodes repair with rate $\Gamma_- = (\Gamma_0/R) \exp{(-f_i \gamma_-)}$. These rates depend on the local frailty $f_i = \sum_{j\in \mathcal{N}(i)} d_j/k_i$, which is the proportion of damaged neighbors of node $i$. This $f_i$ quantifies local damage within the network. Transitions between the damaged and healthy states of nodes are implemented exactly using a stochastic simulation algorithm \cite{Gillespie:1977, Gibson:2000}. For each step, the algorithm chooses a transition to perform from all of the possible transitions. The probability of choosing a particular transition is determined by its transition rate, and after the transition is performed time is incremented by sampling a time increment from an exponential waiting-time distribution with mean rate given by the transition rate. Individual mortality occurs when the two highest degree nodes are both damaged. 

We generate our default network ``topology'' using a linearly-shifted preferential attachment algorithm \cite{Krapivsky:2001,Fotouhi:2013}, which is a generalization of the original Barab\'asi-Albert algorithm \cite{Barabasi:1999}. This generates a scale-free network $P(k) \sim k^{-\alpha}$, where the exponent $\alpha$ and average degree $\langle k \rangle$ can be tuned. (The minimum degree varies as $k_{\mathrm{min}} = \langle k \rangle/2$.) This network is highly heterogeneous in both degree $k_i$ and nearest-neighbor degree  (nn-degree) $k_{i,\mathrm{nn}} = \sum_{j\in \mathcal{N}(i)} k_j/k_i$.   

Since we are concerned with the properties of individual nodes and groups of nodes, we use the same randomly generated network for all individuals. As a result, connections between any two nodes are the same for every individual. To ensure that our randomly generated network is generic, we then redo all of our analysis for $10$ different randomly generated networks. All of these networks behave qualitatively the same, and so we present results averaged over them. Previously \cite{Farrell:2016}, we generated a distinct network realization for each individual. 

We have used observational data for mortality rate and FI vs age to fine-tune the network parameters \cite{Taneja:2016, Farrell:2016}. A systematic exploration of parameters was done in previous work \cite{Taneja:2016, Farrell:2016}.   Most of our parameterization ($N = 10000$, $\alpha=2.27$, $\langle k \rangle=4$, $\gamma_-=6.5$) is the same as reported previously \cite{Farrell:2016}. However, three parameters ($\Gamma_0 = 0.00183/\mathrm{yr}$, $\gamma_+=7.5$, $R = 3$) have been adjusted because we now disallow multiple connections between pairs of nodes during our network generation. This simplifies analysis and adjustment of the network topology, but would also affect mortality rates (see e.g. Fig.~\ref{Mortality Rates} below) without the parameter adjustment. Other network topologies, see Sect.~\ref{network structure}, also use this ``default'' parameterization unless otherwise noted.

Typically, binary deficits have a finite sensitivity \cite{Clegg:2016}, while our model gives us exact knowledge of when a node damages. We have modeled this finite sensitivity by applying non-zero false-negative attribution errors to our raw model FI \cite{Farrell:2016}. This has no effect on the dynamics or on mortality, but does affect the FI. For any raw FI $f_0= \sum_i d_i/n$ from $n$ nodes, there are $n_0 = f_0 n$ damaged nodes. With a false-negative rate of $q$, $n_q$ of these are overturned, where $n_q$ is individually-sampled from a binomial distribution $p(n_q; n_0, 1 - q) =  \binom{n_0}{n_q}(1-q)^{n_q} q^{n_0-n_q}$. We use $f=n_q/n$ as the corrected individual FI. Since our model $f_0$ tends to reach the arithmetic maximum of $1$ at old ages, this effectively gives a maximum observed FI of $\langle f_{\mathrm{max}} \rangle = 1 - q$ \cite{Farrell:2016}. We use $q=0.4$ throughout.

\subsection{Observational Data analysis} 
Observational data is typically ``censored'', meaning that the study ended or an individual dropped out before their death occurred, leaving no specific death age. To avoid this problem, we use a binary mortality outcome e.g. $M = 0$ if an individual is alive within $5$ years of follow-up, or $M = 1$ otherwise. We use $5$ year outcomes throughout for observational data unless otherwise specified. We adapt this approach in our analysis of mutual information \cite{Cover, Shannon:1948}. Our entropy calculations will use binary entropy, $S(M|t) = -p(0|t) \log{p(0|t)} - p(1|t)\log{p(1|t)}$, which we use to calculate information $I(M;D_i|t) = S(M|t) - S(M|D_i,t)$. See also Blokh and Stambler \cite{Blokh:2017}, for other varieties of information analysis for observational data.

We compare our information theory results to a more standard survival analysis with hazard ratios \cite{Spruance:2004}. The hazard ratio is the ratio of instantaneous event rates for two values of an explanatory variable --- e.g. with/without a deficit. A larger hazard ratio means a lower likelihood of surviving with the deficit than without. Hazard ratios are ``semi-parametric'', since they extract the effects of variables on mortality rate from a phenomenological mortality model. We use the Cox proportional hazards model \cite{Cox:1972}, which assumes exponential dependence of mortality rates. We show below that these survival analysis techniques are consistent with our non-parametric mutual information measures.

\subsection{High-$k$ network reconstruction} \label{Reconstruction}
To reconstruct network connections from observed states of nodes, we use the state of each deficit (node) at a given age $t$ (or narrow range of ages in observational data) for each individual in the sample, and calculate the mutual information between individual deficits, $I(D_i;D_j|t)$ \cite{Butte:2000, Margolin:2006}. Connections in the model create correlations between nodes, so a large $I(D_i;D_j|t)$ could indicate a connection. We use data where individuals are the same age (or $\pm$ 5 years in observational data), so that time is not a confounding variable.  Nevertheless, determining whether a given connection exists or not  requires a threshold on $I(D_i;D_j|t)$. If we took this route, we would only assign a connection between nodes if the mutual information is above this threshold.  However, we have no practical way of determining such a threshold, though attempts have been made in the past \cite{Mitnitski:2002}. 

In preliminary tests with our model we have found that matching the reconstructed average degree with the exact average degree is a reliable way of determining a threshold (data not shown), but we still have no way of determining the average degree from observational data. Instead, we use a simple parameter-free method adapted from work on gene co-expression networks \cite{Zhang:2005}. We construct weighted networks, with the mutual information between pairs of nodes as the strength or weight of the connections. We then calculate a ``reconstructed'' degree by adding the information for each possible connection to the node in the network, $\hat{k}_i \equiv \sum_{j\ne i} I(D_i;D_j|t)$ ~\cite{Barrat:2004}. For nodes that aren't connected, $I(D_i;D_j|t) \approx 0$, while $I(D_i;D_j|t)$ is expected to be large for connected nodes. While we cannot reconstruct the actual network, we can reconstruct the rank-order degree of high-$k$ nodes -- since we find that $\hat{k}$ is roughly monotonic with the actual degree $k$ for high-$k$ nodes.

\subsection{Mean-field theory of network dynamics}
\label{MFT}
Here, we present a mean-field theory of our network model to understand the mechanisms underlying our model results. Our mean-field theory (MFT) is based on work on epidemic processes in complex networks by Pastor-Satorras \emph{et al}. \cite{Pastor-Satorras:2015} together with ideas from Gleeson \cite{Gleeson:2011} that we use to include mortality dynamics. 

By MFT we mean a set of deterministic dynamical equations for damage probabilities of network nodes, including mortality nodes. Here, we retain the full degree distribution $P(k)$ and degree correlations $P(k'|k)$ of our stochastic network model. This allows us to identify what model behavior is controlled by the degree distribution and degree correlations. (A simpler MFT, with all nodes having the same degree, has been published \cite{Farrell:2016}.)  With a degree distribution we then solve (see below) thousands of coupled ordinary differential equations (ODEs) with standard numerical integrators.

We average the damaged probabilities $p(d_i=1,t)$ and the undamaged probabilities $p(d_i=0,t)$, conditioned on the damage of the mortality nodes, over all nodes of the same degree $k$:  
\begin{gather*} 
p_{k,d_{m_1},d_{m_2}}(t) \equiv \sum_{\mathrm{deg}(i) = k} p(d_i = 1, d_{m_1},d_{m_2},t)/(NP(k)),\\
q_{k,d_{m_1},d_{m_2}}(t) \equiv \sum_{\mathrm{deg}(i) = k} p(d_i = 0, d_{m_1},d_{m_2},t)/(NP(k)),
\end{gather*}
where the mortality states are indicated by $d_{m_1},d_{m_2} \in \{0,1\}$, $N$ is the number of nodes, and $P(k)$ is the degree distribution. The resulting joint probabilities are $p_{k,d_{m_1},d_{m_2}}$ and $q_{k,d_{m_1},d_{m_2}}$, for damaged and undamaged nodes respectively. These joint probabilities satisfy
\begin{eqnarray}
\sum_{d_{m_1},d_{m_2}} && (p_{k,d_{m_1},d_{m_2}} + q_{k,d_{m_1},d_{m_2}}) = 1, \\
p_{d_{m_1},d_{m_2}} &&= p_{k,d_{m_1},d_{m_2}} + q_{k,d_{m_1},d_{m_2}},\ \textrm{and}\\
p_{k|d_{m_1},d_{m_2}} &&= p_{k,d_{m_1},d_{m_2}}/p_{d_{m_1},d_{m_2}},
\end{eqnarray}
where the first equation is a normalization condition, the second completeness, and the third Bayes' theorem for conditional probabilities. From our mortality rule of $d_{m_1},d_{m_2} = 1$, the probability of mortality is $p_{\mathrm{dead}} =  p_{k,1,1} + q_{k,1,1}$, for any $k$.

The probability of a neighbor of a node of degree $k$ being damaged (which is its local frailty $f$) given a particular mortality state is 
\begin{equation}
	f_{k|d_{m_1},d_{m_2}}(t) = \sum_{k'} P(k'|k) p_{k'|d_{m_1},d_{m_2}},
    \label{local damage}
\end{equation}
where $P(k'|k)$ is the conditional degree distribution, or ``nearest-neighbor'' degree distribution. $P(k'|k)$ describes the structure of connections in the network, and can be varied independently of the degree distribution $P(k)$.   

Writing exact master equations for $N$ nodes is impractical since there would be $2^N$ distinct states to track, with even more distinct transition rates. As an enormous simplification, we use averaged damage and repair rates of nodes of a given connectivity $k$. This is our key mean-field simplification. To do this we approximate $\langle d_id_j\rangle = \langle d_i\rangle\langle d_j\rangle$ for all nodes, and approximate the number of damaged neighbors by a binomial distribution $n_d \sim B(n_d;f_{k| d_{m_1},d_{m_2}}, k) = \binom{k}{n_d}f_{k| d_{m_1},d_{m_2}}^{n_d} (1 - f_{k| d_{m_1},d_{m_2}})^{k - n_d}$ where the average proportion of damaged neighbors will be $f_{k| d_{m_1},d_{m_2}}$. Using Eq.~\ref{local damage}, we can then calculate our MFT damage and repair rates,
\begin{gather} \nonumber
\langle \Gamma_{\pm}(f_{k|d_{m_1},d_{m_2}})\rangle = \Gamma_{0,\pm} \Big\langle \exp{\big(\gamma_{\pm}n_d/k\big)}\Big\rangle\\ 
=\Gamma_{0,\pm} \Big(f_{k|d_{m_1},d_{m_2}}e^{\pm\gamma_{\pm}/k} + 1 - f_{k|d_{m_1},d_{m_2}}\Big)^k. \label{average rates}
\end{gather}
The node degree is explicit in Eq.~\ref{average rates}, while the degree correlation is included through the average local damage in Eq.~\ref{local damage}.

Using these averaged damage/repair rates as transition probabilities, we can write a master equation for nodes with connectivity $k = k_{\mathrm{min}},...,k_{m_2} - 1$ and given the global state of the mortality nodes:
\begin{widetext}
\begin{eqnarray}
\dot{p}_{k,0,0}(t) &=& q_{k,0,0}\langle\Gamma_+(f_k)\rangle - p_{k,0,0}\Big[\langle\Gamma_+(f_{m_1})\rangle + \langle\Gamma_+(f_{m_2})\rangle \Big] 
                 - p_{k,0,0}\langle\Gamma_-(f_k)\rangle + p_{k,1,0}\langle\Gamma_-(f_{m_1})\rangle + p_{k,0,1}\langle\Gamma_-(f_{m_2})\rangle \nonumber \\
\dot{q}_{k,0,0}(t) &=& -q_{k,0,0}\langle \Gamma_+(f_k)\rangle - q_{k,0,0} \Big[\langle\Gamma_+(f_{m_1})\rangle + \langle\Gamma_+(f_{m_2})\rangle \Big] 
                 + p_{k,0,0}\langle\Gamma_-(f_k)\rangle + q_{k,1,0}\langle\Gamma_-(f_{m_1})\rangle + q_{k,0,1}\langle\Gamma_-(f_{m_2})\rangle \nonumber \\
\dot{p}_{k,1,0}(t) &=& q_{k,1,0}\langle\Gamma_+(f_k)\rangle - p_{k,1,0}\langle\Gamma_+(f_{m_2})\rangle + p_{k,0,0}\langle \Gamma_+(f_{m_1})\rangle
                 - p_{k,1,0}\langle\Gamma_-(f_k)\rangle - p_{k,1,0}\langle\Gamma_-(f_{m_1})\rangle \nonumber \\
\dot{q}_{k,1,0}(t) &=& -q_{k,1,0}\langle\Gamma_+(f_k)\rangle - q_{k,1,0}\langle\Gamma_+(f_{m_2})\rangle+ q_{k,0,0}\langle \Gamma_+(f_{m_1})\rangle
                  +q_{k,1,0}\langle\Gamma_-(f_k)\rangle - q_{k,1,0}\langle\Gamma_-(f_{m_1})\rangle \nonumber \\
\dot{p}_{k,0,1}(t) &=& q_{k,0,1}\langle\Gamma_+(f_k)\rangle - p_{k,0,1}\langle\Gamma_+(f_{m_1})\rangle + p_{k,0,0}\langle \Gamma_+(f_{m_2})\rangle
                  - p_{k,0,1}\langle\Gamma_-(f_k)\rangle - p_{k,0,1}\langle\Gamma_-(f_{m_2})\rangle \nonumber \\
\dot{q}_{k,0,1}(t) &=& -q_{k,0,1}\langle\Gamma_+(f_k)\rangle - q_{k,0,1}\langle\Gamma_+(f_{m_1})\rangle+ q_{k,0,0}\langle \Gamma_+(f_{m_2})\rangle 
                  +p_{k,0,1}\langle\Gamma_-(f_k)\rangle - q_{k,0,1}\langle\Gamma_-(f_{m_2})\rangle \nonumber \\
\dot{p}_{k,1,1}(t) &=& p_{k,1,0}\langle\Gamma_+(f_{m_2})\rangle+ p_{k,0,1}\langle \Gamma_+(f_{m_1})\rangle \nonumber \\
\dot{q}_{k,1,1}(t) &=& q_{k,1,0}\langle\Gamma_+(f_{m_2})\rangle+ q_{k,0,1}\langle \Gamma_+(f_{m_1})\rangle.
\label{MFTequations}
\end{eqnarray}
\end{widetext}
In these equations we have not shown the mortality state indices of $f_k$ for readability, but they are the same as the associated $p$ or $q$ factors. We have also defined $f_{m_1}$ and $f_{m_2}$ as the local frailties of the first and second mortality node, respectively. We have $8$ equations for each distinct degree $k$. The last two equations determine the mortality rate, $ \dot{p}_{k,1,1}+\dot{q}_{k,1,1}$.

The mean-field model couples the dynamics of the lowest degree ($k=2$) with all degrees up to the two highest  (mortality nodes). Solving the equations requires us to explicitly determine the two mortality node degrees. While approximate calculations of the maximum degree of scale-free networks are available \cite{Dorogovtsev:2002}, we need the two highest degrees. We use $k_{m_1} = 885$ and $k_{m_2} = 768$, based on the averages from simulations of the network. Similarly, we use $k_{m_1} = 14$ and $k_{m_2} = 13$ for ER random networks and $k_{m_1} = 7$ and $k_{m_2} = 6$ for WS small-world networks.  Qualitatively, our qualitative MFT results do not depend on these mortality node degrees, as long as they are sufficiently large. The minimum degree $k_{\mathrm{min}}$ is determined by the network topology.

Our default model uses a linearly-shifted preferential-attachment model, which has explicit functional forms for the degree distribution $P(k)$ and the nearest-neighbor degree distribution $P(k'|k)$ as $N \to \infty$ \cite{Fotouhi:2013}.

We numerically solve Eq.~\ref{MFTequations} for the probabilities $p_{k,d_{m_1},d_{m_2}}(t)$ and $q_{k,d_{m_1},d_{m_2}}(t)$. These then allow us to calculate the average FI,
\begin{eqnarray} \label{MFT FI}
\langle F(t)\rangle &=& 
\frac{\sum\limits_{k = k_{\mathrm{low}}}^{k_{\mathrm{high}}} P(k) p_{k|\mathrm{alive}}}
{\sum\limits_{k = k_{\mathrm{low}}}^{k_{\mathrm{high}}} P(k)},\\
p_{k|\mathrm{alive}} &\equiv&
	\frac{p_{k,0,0} + p_{k,0,1} + p_{k,1,0}}
    	{p_{k,0,0} + p_{k,0,1} + p_{k,1,0} + q_{k,0,0} + q_{k,0,1} + q_{k,1,0}},
        \nonumber
\end{eqnarray} 
so that the average is over the surviving individuals. Our averaged damage-rates overestimate the true values, so for the same parameterization mortality occurs on a shorter timescale in the MFT. This is because rapidly damaging nodes drop out of the full model once they are damaged, but continue to contribute to the average damage rates in the mean-field model through Eq.~\ref{average rates}.  Because of this, when plotting MFT results we scale time by $t_{\mathrm{scale}}$, the time at which every node is damaged ($p_{k} = 1$).

\section{Results} 
We will focus on measures that can be compared between model and observational data, or that provide insight into the network structure of organismal aging. We start with observational data, to expand the observed aging phenomenology. Then we explore how our network model behaves, with a focus on how network properties determine the qualitative behavior of the model. 

\subsection{Observational Data} 
Dauntingly, we have three challenges for assessing network properties from observational data: human studies are small (typically with $\lesssim 10^4$ individuals) so that results will be noisy, different studies will have quantitative differences due to cohort differences and choices of measured health attributes, and we have no robust way of reconstructing networks from observed deficits so that the absolute connectivity of health-attributes is unknown. We face these challenges by focusing on qualitatively robust behavior from larger observational studies; this will also help us to confront our results with the behavior of our generic network model.

From the American National Health and Nutrition Examination Survey (NHANES, see \cite{NHANES:2014}), the 2003-2004 and 2005-2006 cohorts were combined, with up to $5$ years of mortality reporting (one measurement of age and FI with either age of death or last age known to be still alive). Laboratory data were available for $9052$ individuals and clinical data on $10004$, aged $20+$ years. Thresholds used to binarize lab deficits are found in \cite{Blodgett:2017}. From the Canadian Study of Health and Aging (CSHA, see \cite{CSHA:1994}), $5$ year mortality reporting are obtained from 1996/1997. Laboratory data were available for $1013$ individuals and clinical data for $8547$, aged $65+$ years. Thresholds used to binarize lab deficits are found in \cite{Howlett:2014}. By approaching both the NHANES and CSHA studies with the same approaches, we can identify qualitatively robust features of both. 

\begin{figure} 
  \begin{minipage}[htb]{0.45\textwidth}
    \includegraphics[trim=3mm 5mm 5mm 3mm, clip,width=\textwidth]{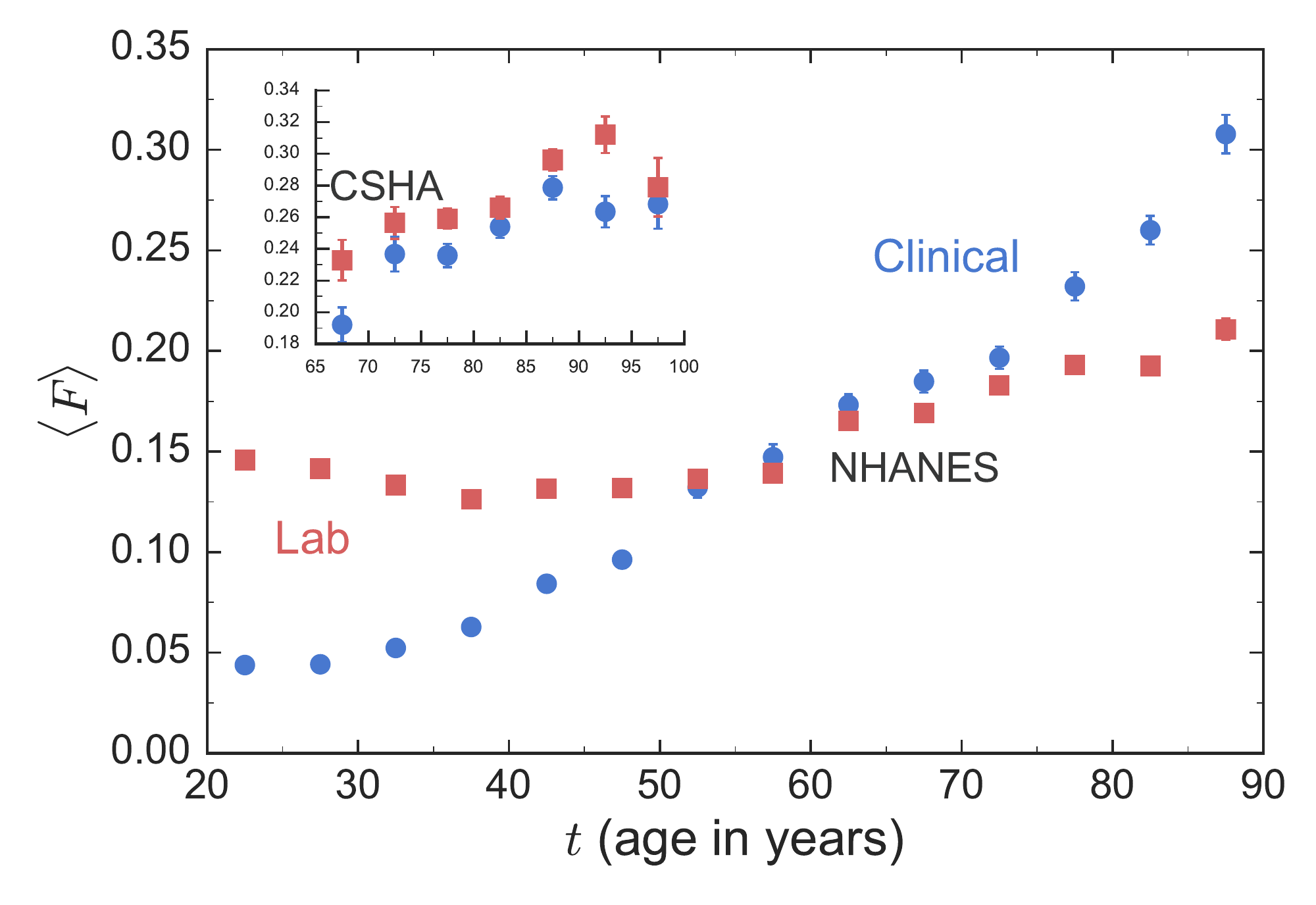}
    \caption{Average FI vs age $t$ with $\langle F_{\mathrm{lab}} \rangle$ (red squares) and $\langle F_{\mathrm{clin}} \rangle$ (blue circles) from the NHANES dataset (main figure). The inset shows the same plot for the CSHA dataset. Error bars show the standard error of the mean. All individuals used in this plot have both $F_{\mathrm{clin}}$ and $F_{\mathrm{lab}}$ measured.}
    \label{Observational Time Structure}
  \end{minipage}
\end{figure}

Fig.~\ref{Observational Time Structure} shows the average FI vs age for $F_{\mathrm{lab}}$ in red and $F_{\mathrm{clin}}$ in blue for the NHANES in the main plot and CSHA in the inset. In both studies lab deficits accumulate earlier than clinical deficits. A crossover appears in the NHANES data around age $55$ after which clinical deficits are more damaged than lab deficits. A similar crossover does not appear to happen in the CSHA data. 

\begin{figure} 
  \begin{minipage}[htb]{0.45\textwidth}
    \includegraphics[trim=6mm 10mm 5mm 3mm, clip,width=\textwidth]{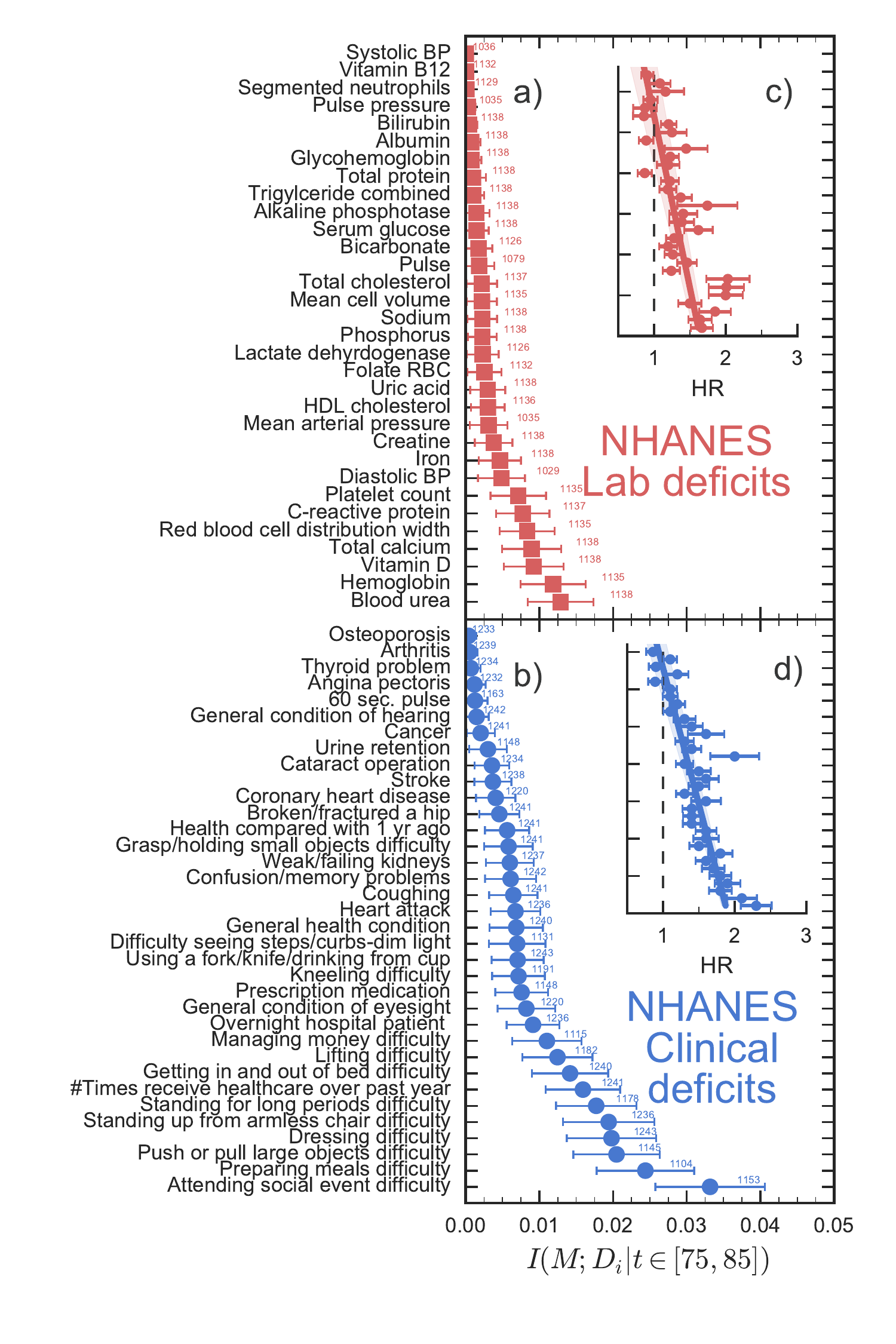}
    \caption{Rank-ordered deficits in terms of information $I(M;D_i|t\in[75,85])$ for the NHANES dataset. Red points are lab deficits, blue points are clinical deficits. Error bars are standard errors found from bootstrap re-sampling. Small numbers next to the points indicate the number of individuals that were available in the data for the corresponding deficit. Insets show the corresponding hazard ratios for the deficits found from a Cox proportional hazards model regression, with the deficit and age used as covariates. The error bars show standard errors, and the line shows a linear regression through these points with the standard error in slope and intercept shown in a lighter color.}
    \label{NHANES Information Fingerprint}
  \end{minipage}
\end{figure}

\begin{figure} 
  \begin{minipage}[htb]{0.45\textwidth}
    \includegraphics[trim=10mm 10mm 5mm 3mm, clip,width=\textwidth]{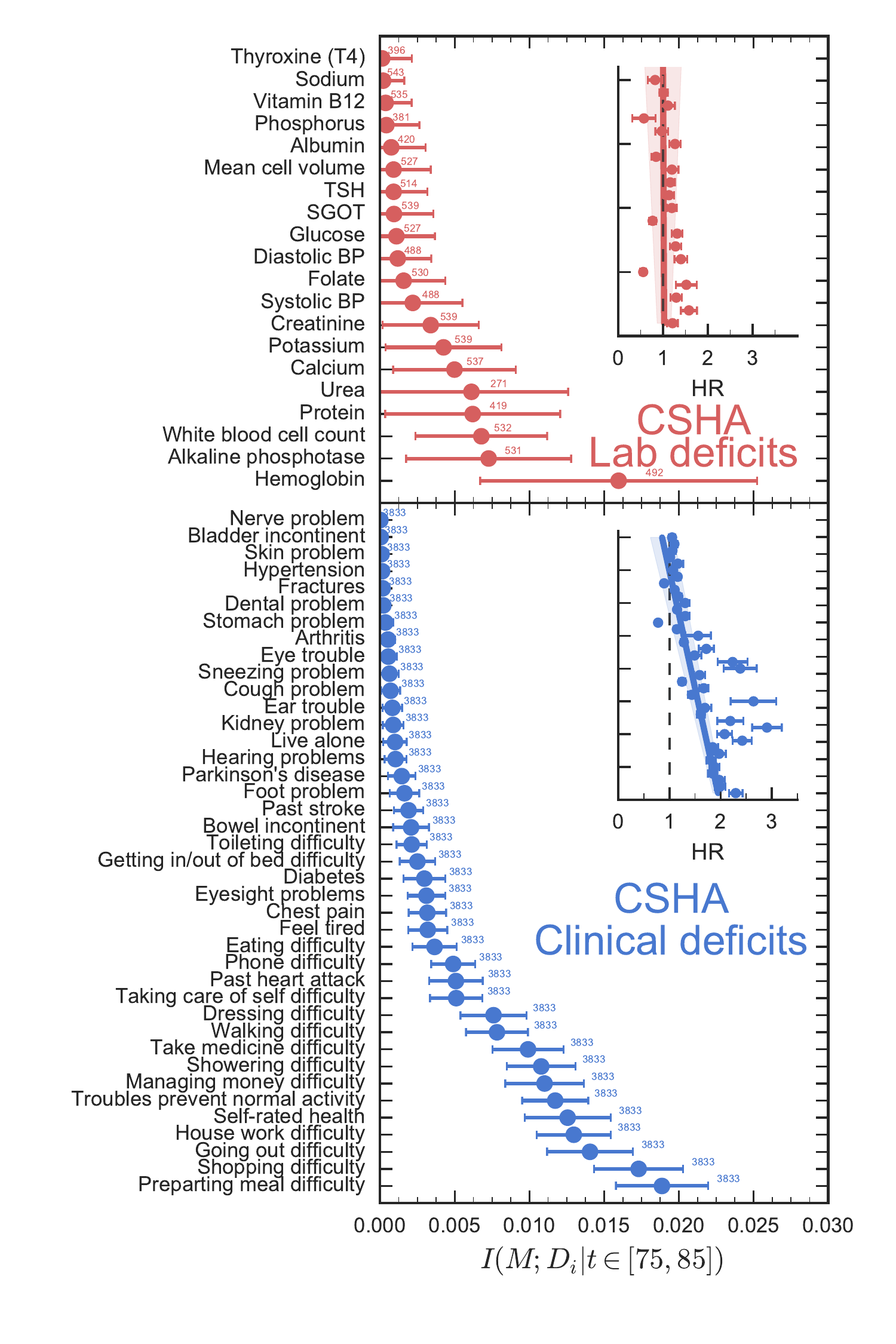}
    \caption{Rank-ordered deficits in terms of information $I(M;D_i|t\in[75,85])$ for the CSHA dataset. Red points are lab deficits, blue points are clinical deficits. Error bars are standard errors found from bootstrap re-sampling. Small numbers next to the points indicate the number of individuals that were available in the data for the corresponding deficit. Insets show the corresponding hazard ratios for the deficits found from a Cox proportional hazards model regression, with the deficit and age used as covariates. The error bars show standard errors, and the line shows a linear regression through these points with the standard error in slope and intercept shown in a lighter color.}
    \label{CSHA Information Fingerprint}
  \end{minipage}
\end{figure}

Figs.~\ref{NHANES Information Fingerprint} and \ref{CSHA Information Fingerprint}  show deficits rank-ordered in information $I(M;D_i|t)$ for the NHANES and CSHA studies, respectively. These are  information ``fingerprints''. Red points correspond to lab deficits and blue to clinical deficits, as indicated. Both types of deficits have similar magnitudes of information, although clinical deficits are typically more informative. The comparable magnitudes of mutual information for the majority of individual deficits between lab and clinical FIs is consistent with earlier analysis that found similar association between lab and clinical FIs with mortality using survival analysis \cite{Howlett:2014, Blodgett:2017, Blodgett:2016}.

Insets in Figs.~\ref{NHANES Information Fingerprint} and \ref{CSHA Information Fingerprint} show the corresponding hazard ratio (HR) for the deficit found from a Cox proportional hazards model regression, with the deficit value and age used as covariates. This semi-parametric analysis is often done with medical data \cite{Jones:2005}. The HR tends to increase as the rank-ordered information increases, indicating that our mutual-information approach is capturing similar effects. Nevertheless, we prefer mutual-information because it is non-parametric (not model-dependent) and so relies on fewer assumptions. 

Our deficit-level analysis highlights the great variability of mutual information (and HR ratios) between individual deficits.  We have  shown that lab and clinical deficits have a range of mutual information. We further note that the top 5 - 7 most informative clinical deficits in both the NHANES and CSHA datasets  measure functional disabilities or dysfunction \cite{Rockwood:2017b}. We find that these high level deficits are the most informative of mortality, and more informative than any of the lab deficits.  From this, we hypothesize that highly informative clinical deficits will also be highly connected. 

\begin{figure} 
  \begin{minipage}[thb]{0.45\textwidth}
    \includegraphics[trim=5mm 6mm 3mm 3mm, clip,width=\textwidth]{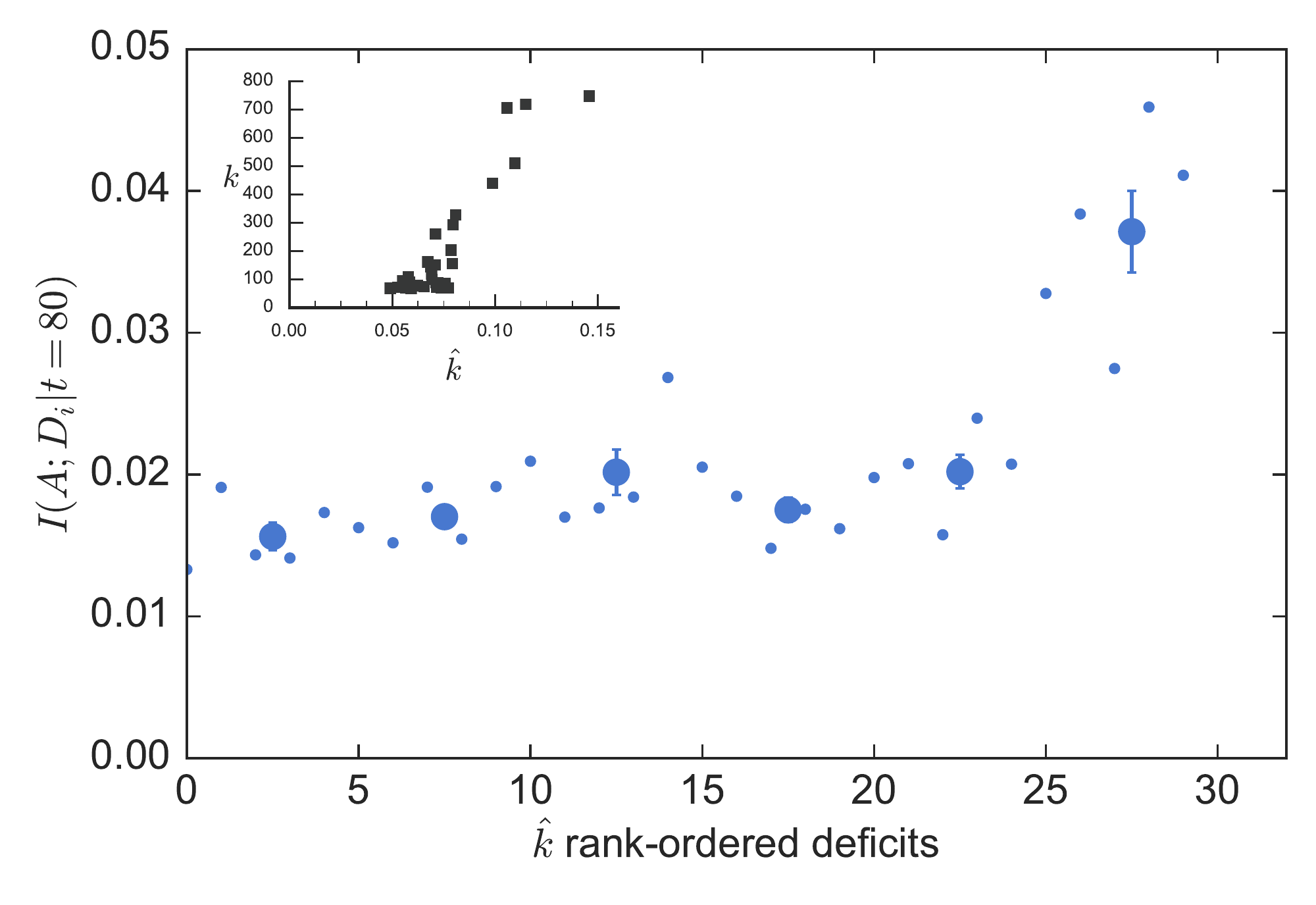}
    \caption{Information $I(A;D_i|t=80)$ vs rank-ordered deficits using reconstructed degree $\hat{k}_i$, for our computational model. The top $32$ most connected nodes are reconstructed with $10000$ individuals. The smaller (blue circles) points show individual nodes, the larger points show a binned average, and error bars are the standard error of the mean within each bin. The inset (black squares) shows the exact degree $k$ vs the reconstructed degree $\hat{k}$.}
    \label{Reconstruction Validation}
  \end{minipage}
\end{figure}

\begin{figure} 
  \begin{minipage}[thb]{0.45\textwidth}
    \includegraphics[trim=10mm 10mm 5mm 5mm, clip,width=\textwidth]{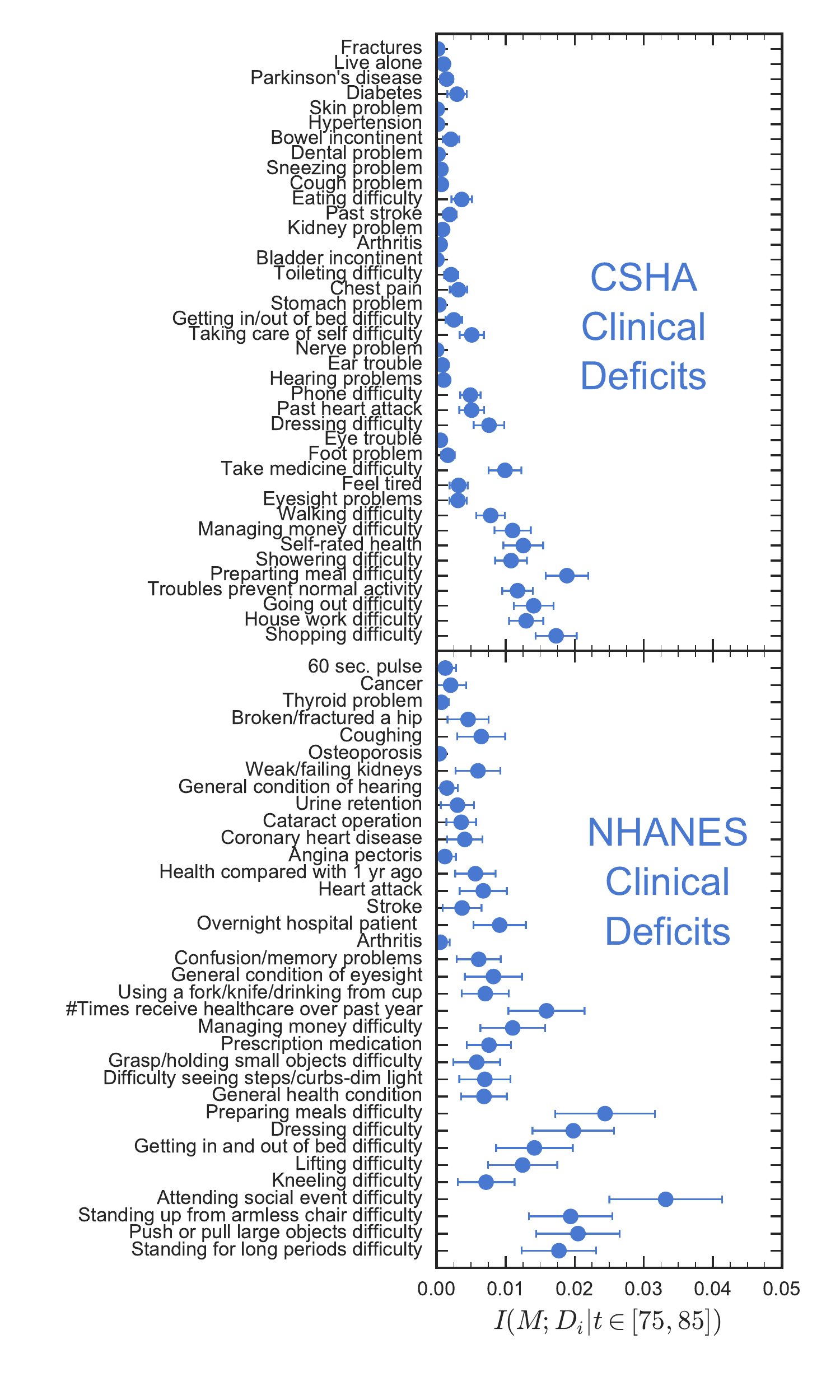}
    \caption{Rank-ordered clinical deficits in terms of reconstructed degree $\hat{k}$ vs information with respect to mortality $I(M;D_i|t\in[75,85])$ for the NHANES and CSHA datasets. The reconstruction algorithm is detailed in Sec. \ref{Reconstruction}. Error bars are standard errors found from bootstrap re-sampling. }
    \label{Reconstructed Clinical Fingerprint}
  \end{minipage}
\end{figure}

We have been able to partially reconstruct the network structure of clinical measures, as detailed in Sec. \ref{Reconstruction}. In Fig.~\ref{Reconstruction Validation}, we have validated this approach with the top $32$ most-connected model nodes. We use $10000$ individuals for our validation, approximately the same number of people we have available in the observational studies. We know that our model information tends to increase with degree for the high degree nodes (see \cite{Farrell:2016}, and also Fig.~\ref{Model Spectra} below). Fig.~\ref{Reconstruction Validation} shows that information also increases with the reconstructed degree $\hat{k}$, as expected for a good reconstruction. The inset showing $k$ vs $\hat{k}$ indeed shows that the reconstructed degree is approximately monotonic with the exact degree --- especially at higher $k$. 

This means the reconstructed degree should provide a reasonable rank-order in connectivity for observational data. Nevertheless, low-degree nodes are not reliably rank-ordered. Accordingly we only attempt to reconstruct clinical $\hat{k}$ with this approach.

In Fig.~\ref{Reconstructed Clinical Fingerprint}, we plot information with respect to mortality $I(M;D_i|t\in[75,85])$ for each deficit, where deficits are rank-ordered in terms of reconstructed degree $\hat{k}$. Information increases with reconstructed degree for both the NHANES and CSHA clinical data. This shows that high information deficits correspond to  high connectivity in the observational data. Also, nearly all of the functional disabilities intuitively hypothesized to have a high connectivity are also found to have a large reconstructed degree.

\subsection{Model Age-structure} 
We saw, in Fig.~\ref{Observational Time Structure}, that pre-clinical (lab) damage accumulates before clinical damage in observational data. This is a qualitatively robust observation, seen in both NHANES and CSHA observational data. We also observed, in Fig.~\ref{Reconstructed Clinical Fingerprint}, that (in terms of rank order) highly connected clinical deficits were more informative than less connected deficits. We expect that health-attributes assessed by laboratory tests are less connected than the high level functional attributes assessed clinically. We hypothesize that $F_{\mathrm{lab}}$ and $F_{\mathrm{clin}}$ should behave \textit{qualitatively} like collections of low or high degree nodes, respectively, within our network model of aging. 

We construct two distinct FIs to capture the difference between well-connected hub nodes and poorly connected peripheral nodes. We measure low-degree damage by constructing $F_{\mathrm{low}}=\sum_i d_i/n$ from a random selection of $n=32$ nodes all with $k = k_{\mathrm{min}}  = 2$. Similarly, we measure high-degree damage with $F_{\mathrm{high}}$ from the top $32$ most connected nodes (excluding the two most connected nodes, which are the mortality nodes).

\begin{figure} 
  \begin{minipage}[thb]{0.45\textwidth}
    \includegraphics[trim=5mm 5mm 2mm 2mm, clip,width=\textwidth]{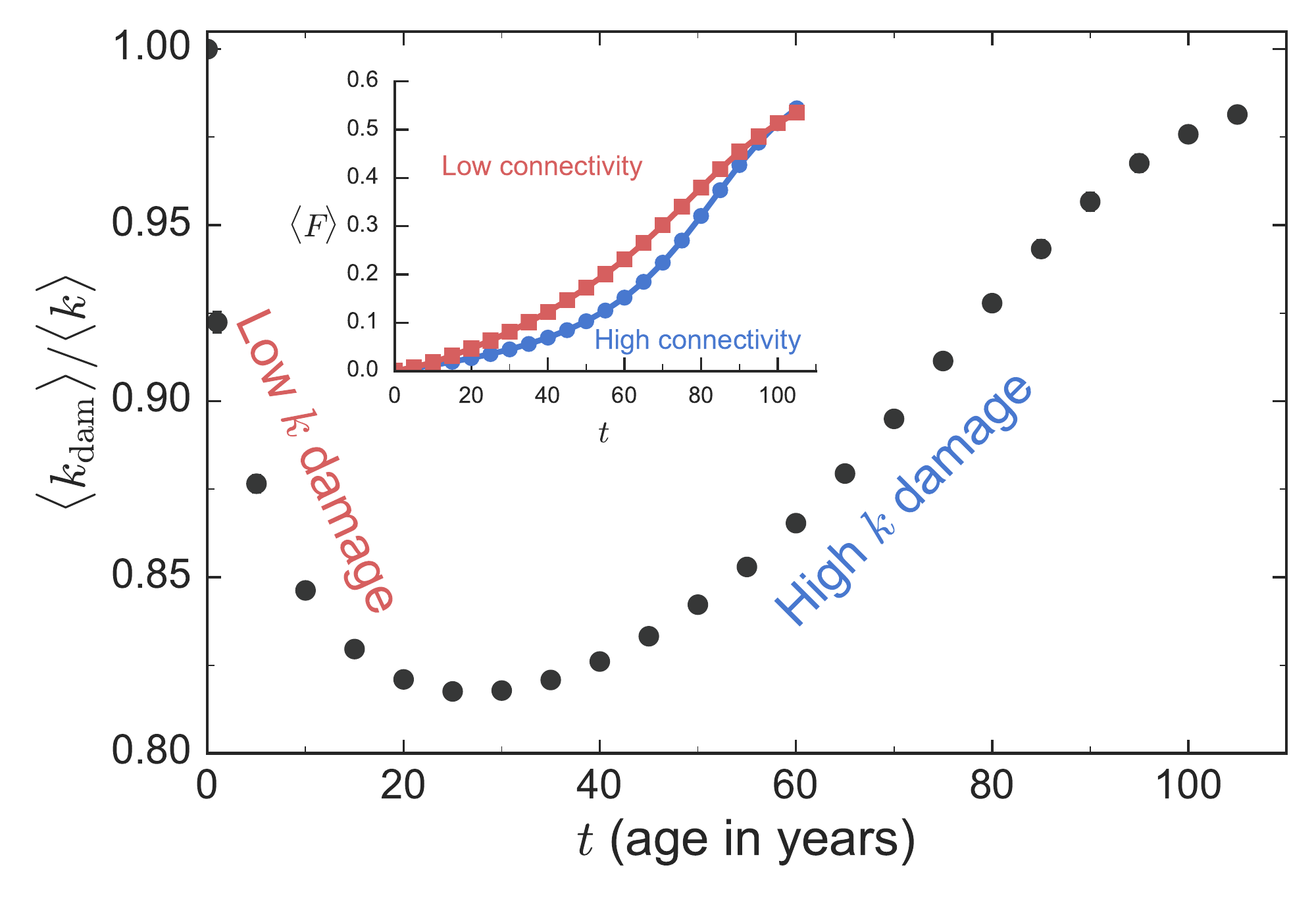}
    \caption{Average degree of damaged model deficits $\langle k_{\mathrm{dam}}(t) \rangle$, scaled by the average degree of the network $\langle k \rangle$, vs time $t$. Error bars, barely visible at low $t$, represent the standard deviation between randomly generated networks.  As indicated, at earlier times low-connectivity nodes are preferentially damaged while at later times higher connectivity nodes are preferentially damaged. The inset shows the average damage of low-connectivity nodes $\langle F_{\mathrm{low}}\rangle$ (red squares) and of high-connectivity nodes $\langle F_{\mathrm{high}}\rangle$ (blue circles) vs age.}
    \label{Model Time Structure}
  \end{minipage}
\end{figure}

Fig.~\ref{Model Time Structure} shows the cumulative average degree of damaged nodes $\langle k_{\mathrm{dam}} \rangle = \langle \sum_{i = 0}^N k_i d_i / \sum_{i = 0}^N d_i \rangle$ vs age $t$. Error bars represent the standard deviation between 10 different randomly generated networks. They are each comparable to or smaller than the point size, indicating that the age-structure represents the network topology rather than a single network realization.  

For a uniform network or for damage rates independent of the degree of a node, we would expect $\langle k_{\mathrm{dam}} \rangle = \langle k \rangle$ for all ages $t$. However, we see the average degree of damaged deficits start at $\langle k \rangle$, with an initial decrease until around age $25$ and then an increase back to $\langle k \rangle$ --- implying damage does not uniformly propagate through the network.  

Initially damage is purely random, so $\langle k_{\mathrm{dam}}(0) \rangle = \langle k \rangle$. Nodes with degree $k_i < \langle k \rangle$ are being damaged when $\langle k_{\mathrm{dam}} \rangle/\langle k \rangle$ decreases from $1$, and nodes of degree $k_i > \langle k \rangle$ are being damaged when $\langle k_{\mathrm{dam}} \rangle/\langle k \rangle$ increases towards $1$. 

The inset of Fig.~\ref{Model Time Structure} shows the average FI vs age for $F_{\mathrm{low}}$ and $F_{\mathrm{high}}$. We see $\langle F_{\mathrm{low}}\rangle$ initially larger than $\langle F_{\mathrm{high}}\rangle$. Eventually with age, $\langle F_{\mathrm{high}}\rangle$ increases to match $\langle F_{\mathrm{low}}\rangle$ and even slightly exceed at very old ages. Thus, low-$k$ nodes behave similarly to lab deficits, and high-$k$ nodes behave similarly to clinical deficits in observational health data. Low-$k$ nodes and lab measures both damage early and high-$k$ nodes and clinical measures both damage late.

We have not tuned our model parameterization to obtain this age-structure of damage in the network model. Indeed, for other parameter choices we see qualitatively similar behavior (data not shown) for the scale-free networks that we have been using. To better understand this age-structure we consider the effects of network connectivity within our mean-field theory.

\begin{figure} 
  \begin{minipage}[th]{0.45\textwidth}
    \includegraphics[trim=5mm 5mm 3mm 3mm, clip, width=\textwidth]{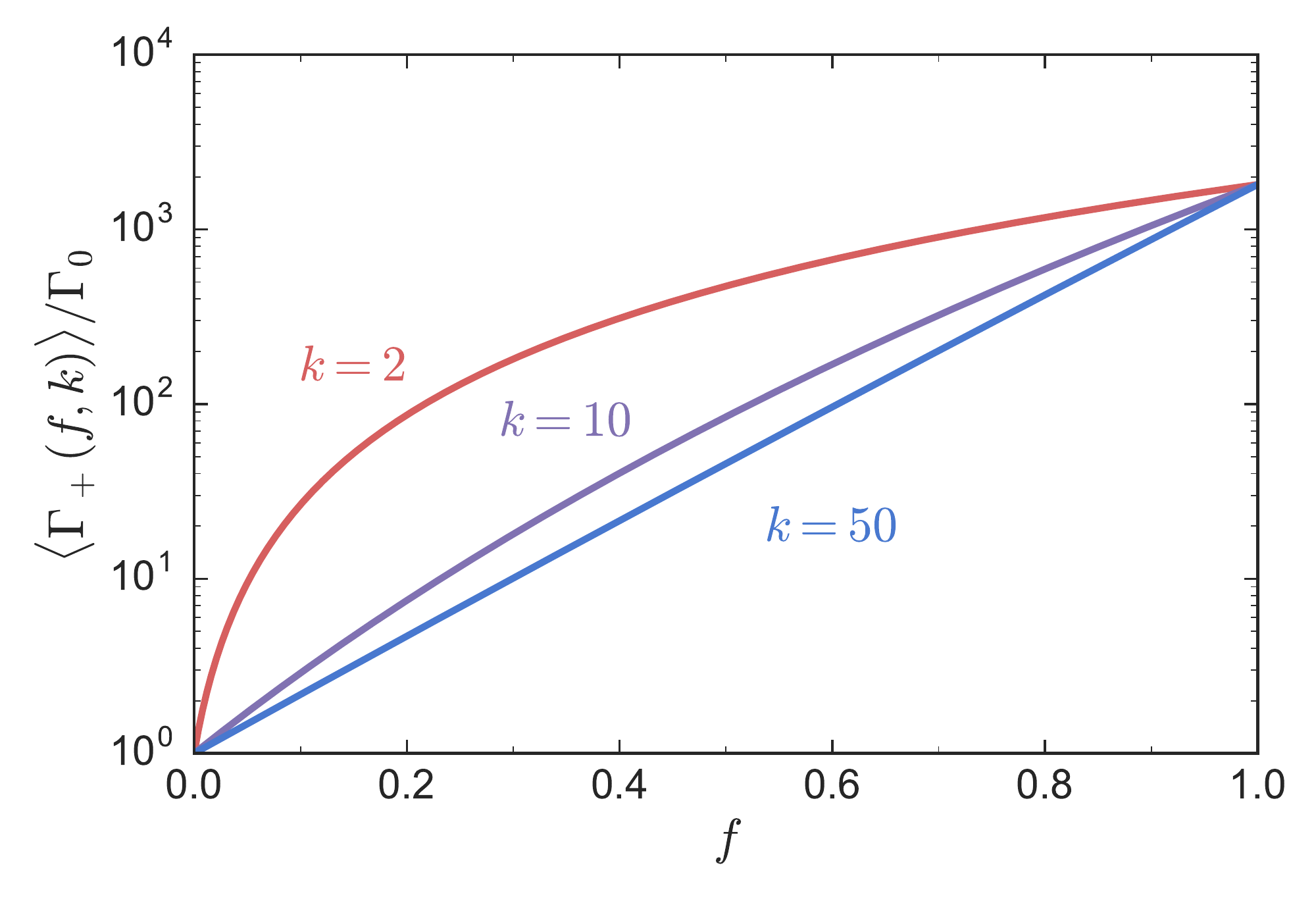}
    \caption{Average mean-field damage rates $\langle \Gamma_+\rangle/\Gamma_0$ for nodes of a given degree $k$ (as indicated) vs the local frailty of these nodes $f$, as given by Eq.~\ref{average rates}. Low-connectivity nodes exhibit significantly higher damage rates at intermediate values of $f$.}
    \label{Damage Rates Figure}
  \end{minipage}
\end{figure}

In our mean-field theory, we find our averaged damage rates explicitly depend on $k$ in Eq.~\ref{average rates}. This is shown in Fig.~\ref{Damage Rates Figure}, these mean-field damage rates increase with smaller $k$ at a given $f$. This results from Jensen's inequality, since the damage rate is convex in the local frailty $f$ and the lower degree nodes will have a broader distribution of local frailty for the same global frailty. This implies that low-$k$ nodes should damage more frequently until they are exhausted and $F_{\mathrm{low}}$ saturates. 

We can confirm this with the full MFT results. We can determine the FI from Eq.~\ref{MFT FI} and  calculate both  $F_{\mathrm{high}}$ and $F_{\mathrm{low}}$ by choosing which degrees to include. The $k_{\mathrm{low}}$ and $k_{\mathrm{high}}$ determine the nodes included in the FI.  For $F_{\mathrm{high}}$, we choose $k_{\mathrm{high}} = k_{m_2} - 1$ and $k_{\mathrm{low}}$ so that $N\sum_{k=k_{\mathrm{low}}}^{k_{\mathrm{high}}}P(k) = n \simeq 32$ for the smallest possible $k_{\mathrm{low}}$ (32 is the number of FI nodes typically used in our model and observational studies). For $F_{\mathrm{low}}$, we choose $k_{\mathrm{low}}=k_{\mathrm{min}}$ and choose the smallest $k_{\mathrm{high}}$ so that $n \simeq 32$. 

We also calculate
\begin{equation}
\langle k_{\mathrm{dam}}(t) \rangle =\frac{\sum\limits_{k = k_{\mathrm{low}}}^{k_{\mathrm{high}}} kP(k) p_{k|\mathrm{alive}}}{\sum\limits_{k = k_{\mathrm{low}}}^{k_{\mathrm{high}}} P(k)p_{k|\mathrm{alive}}},
\end{equation}
which is the cumulative average degree of damaged nodes as was done for our computational results.

\begin{figure} 
  \begin{minipage}[th]{0.45\textwidth}
    \includegraphics[trim=5mm 5mm 3mm 3mm, clip, width=\textwidth]{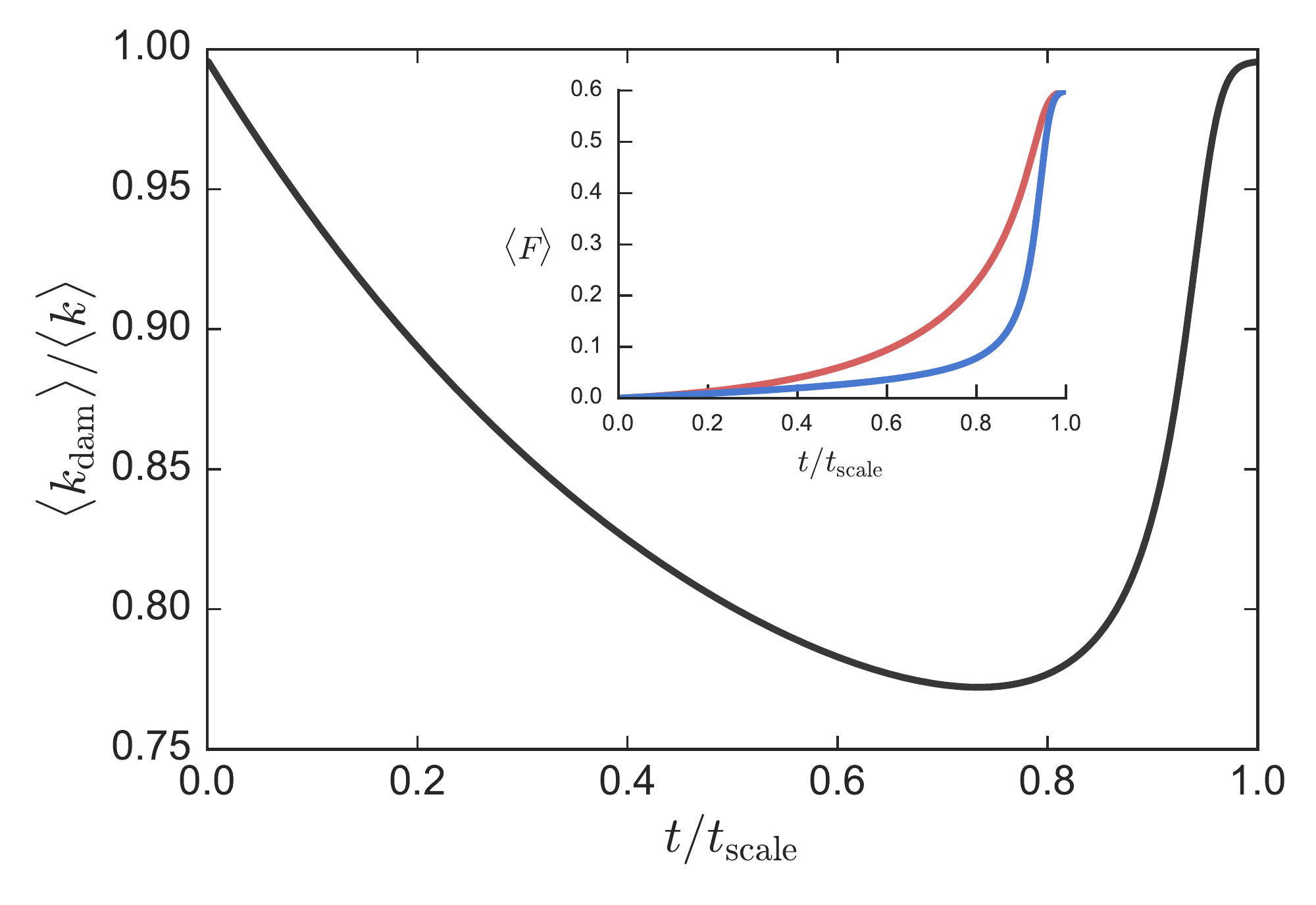}
    \caption{From our mean-field calculation in Sec. ~\ref{MFT}, we show the average degree of damaged deficits $\langle k_{\mathrm{dam}} \rangle$ scaled by the average network degree $\langle k \rangle$ vs time scaled by the time when the network becomes fully damaged, $t/t_{\mathrm{scale}}$. The inset shows the average damage of high connectivity nodes $\langle F_{\mathrm{high}}\rangle$ in blue and low connectivity nodes $\langle F_{\mathrm{low}}\rangle$ vs the scaled time.}
    \label{Mean Field Time Structure}
  \end{minipage}
\end{figure}

In Fig.~\ref{Mean Field Time Structure} the age-structure from the mean-field calculation shows the same early damage of low-$k$ nodes shown in Fig.~\ref{Model Time Structure} and (in inset) the more-rapid growth of $F_{\mathrm{low}}$ compared to $F_{\mathrm{high}}$ at earlier times. Our mean-field calculation also shows a more-rapid growth of $F_{\mathrm{high}}$ compared to $F_{\mathrm{low}}$ at later times, as shown in the inset of Fig.~\ref{Mean Field Time Structure}.  This largely is explained by the saturation of $F_{\mathrm{low}}$.

We conclude that the age-structure seen observationally and in our network model, can be explained by the degree distribution and neighbor-degree correlations of our MFT.  This motivates us to investigate how node degree and neighbor-degree affect mortality within the context of our network model.

\begin{figure} 
  \begin{minipage}[th]{0.45\textwidth}
    \includegraphics[trim=14mm 16mm 6mm 6mm, clip,width=\textwidth]{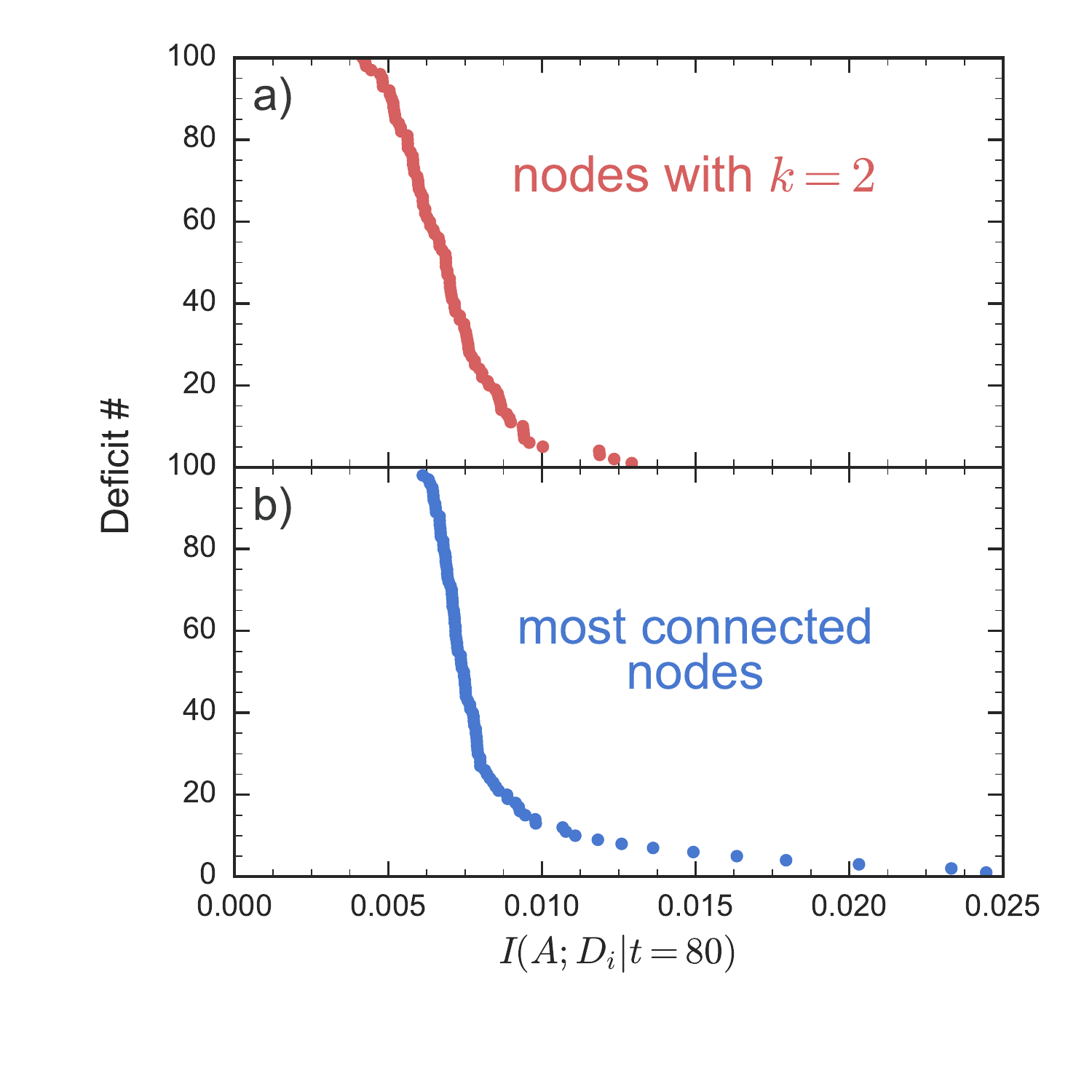}
    \caption{Mutual-information of selected model deficits $I(A;D_i| t = 80)$ at age $80$ years, averaged over $10$ randomly generated networks with $10^7$ individuals each and rank-ordered. Red points are low-$k$ deficits, blue points are high-$k$ deficits.} 
    \label{Model Information Fingerprint}
  \end{minipage}
\end{figure}

\subsection{Model Node Information} 
Fig.~\ref{Model Information Fingerprint} shows the mutual information between death age and individual nodes $I(A;D_i)$ for our model. Red points are a random selection of $100$ low-connectivity nodes all with $k = k_{\mathrm{min}} = 2$, the blue points are the top $100$ most connected nodes (excluding the $2$ mortality nodes). For each selection, we have rank-ordered the nodes in terms of mutual-information. The mutual-information for both high and low connectivity nodes are comparable. This is surprising since previous work showed a monotonic increase of the average information with connectivity \cite{Farrell:2016}.   However that work used a different network for each individual, so that network properties other than the average degree were lost by pooling nodes of the same degree. 

Without parameter tuning, we obtain striking qualitative agreement of the magnitude of the mutual-information with mortality for both model and observational data (see Figs.~\ref{NHANES Information Fingerprint} and \ref{CSHA Information Fingerprint}). We also obtain an overlap of magnitudes of the mutual-information of low-degree and high-degree nodes that is similar to that seen between pre-clinical and clinical deficits.  Since we know the model network connectivity, we can now examine what network properties cause this behavior for our model.

\begin{figure} 
  \begin{minipage}[htb]{0.45\textwidth}
    \includegraphics[trim=17mm 16mm 6mm 6mm, clip,width=\textwidth]{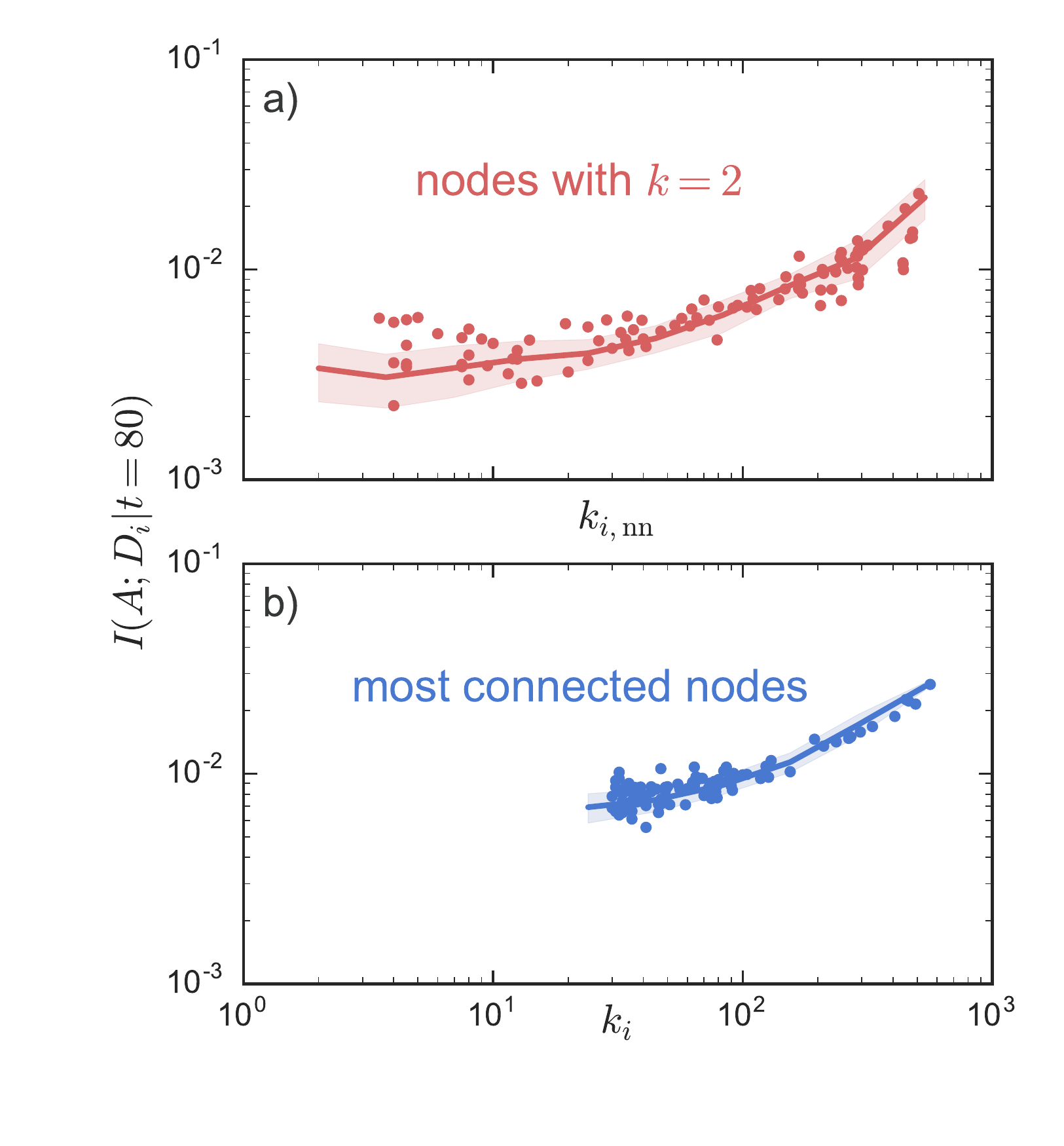}
    \caption{Model information spectra $I(A;D_i|t=80)$ vs degree $k_i$ for the top $100$ most connected nodes in blue, or vs $k_{i,\mathrm{nn}}$ for a random selection of $100$ peripheral nodes all with $k=k_{\mathrm{min}} =2$. Points show a sample of a single network, line shows an average over $10$ randomly generated networks and the random choice of $100$ nodes with $k=2$, the shaded error region shows the standard deviation over the random networks.}
    \label{Model Spectra}
  \end{minipage}
\end{figure}

In Fig.~\ref{Model Spectra}, we show the ``spectrum'' of mutual information between death age and  individual nodes $I(A;D_i|t=80)$. We use individuals at age $t=80$ years, where the mutual information is close to maximal \cite{Farrell:2016}. We use the same network for every individual, so that we do not lose the properties of the network between individuals. For the most connected nodes, in blue, we plot mutual information vs. the connectivity of the nodes.  Here we see the monotonic trend of mutual information vs connectivity, though there is significant variation for individual nodes. For the least connected nodes, in red, all of the nodes have $k=2$. Instead of connectivity, we considered the nearest neighbor degree $k_{i,\mathrm{nn}} = \sum_{j\in \mathcal{N}(i)} k_j/k_i$ --- i.e. the connectivity of  the neighbors of a node. With respect to  $k_{\mathrm{nn}}$, we see a similar monotonic increase of the mutual information for $k=2$ nodes. 

Neighbor-connectivity $k_{\mathrm{nn}}$ is predictive of mortality for minimally connected nodes. We hypothesize that this is because the neighbor-connectivity affects when peripheral ($k=2$) nodes are damaged, i.e. that peripheral nodes with low-$k_{\mathrm{nn}}$ are damaged earlier than those with large $k_{\mathrm{nn}}$.

\begin{figure} 
  \begin{minipage}[htb]{0.45\textwidth}
    \includegraphics[trim=6mm 6mm 3mm 3mm, clip,width=\textwidth]{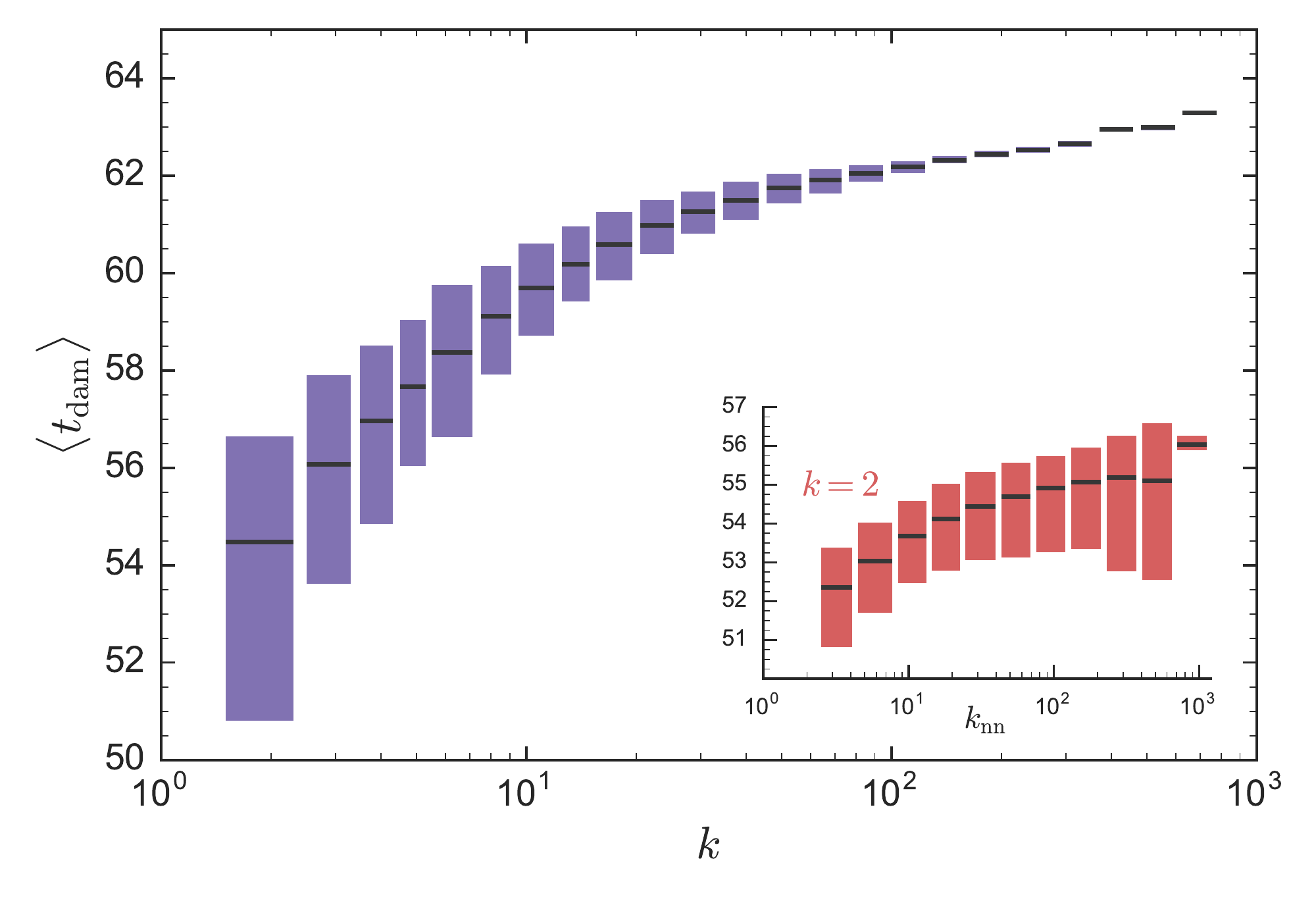}
    \caption{Average time of damage $\langle t_{\mathrm{dam}}\rangle$ vs degree $k$ for all non-mortality nodes in the network. Inset shows $\langle t_{\mathrm{dam}}\rangle$ for $k=2$ nodes vs nn-degree $k_{\mathrm{nn}}$. Nodes are binned based on $k$. The solid colored bars represent the entire range of average damage times observed for individual nodes within a bin, while the horizontal black lines indicate the average over the bin. All results are averaged for $10$ randomly generated networks.}
    \label{Model knn damage}
  \end{minipage}
\end{figure}

In the inset of Fig.~\ref{Model knn damage} we confirm that high-$k_{\mathrm{nn}}$ $k=2$ nodes damage later. This allows high-$k_{\mathrm{nn}}$ nodes to be informative of mortality because they are diagnostic of a more highly damaged network. From Fig.~\ref{Model knn damage} we see that there is a large range of times for which lower-$k$ nodes damage. Nevertheless, on average the  high-$k_{\mathrm{nn}}$ nodes at $k=2$ damage before high-$k$ nodes even though (see Fig.~\ref{Model Spectra}) they can be similarly informative. 

\subsection{Model Network Structure} 
\label{network structure}
We have seen that our network model of aging is able to capture detailed behavior of lab and clinical FIs such as the the larger damage rates for low-$k$ nodes at the same time as the surprising informativeness of some low-$k$ nodes. The network is an important aspect of our model, and so far we have assumed that it is a preferential attachment scale-free network \cite{Krapivsky:2001, Fotouhi:2013, Barabasi:1999}. In this section, we explore the qualitative behavior of different network topologies. 

\begin{figure} 
  \begin{minipage}[htb]{0.45\textwidth}
    \includegraphics[trim=5mm 7mm 5mm 3mm, clip,width=\textwidth]{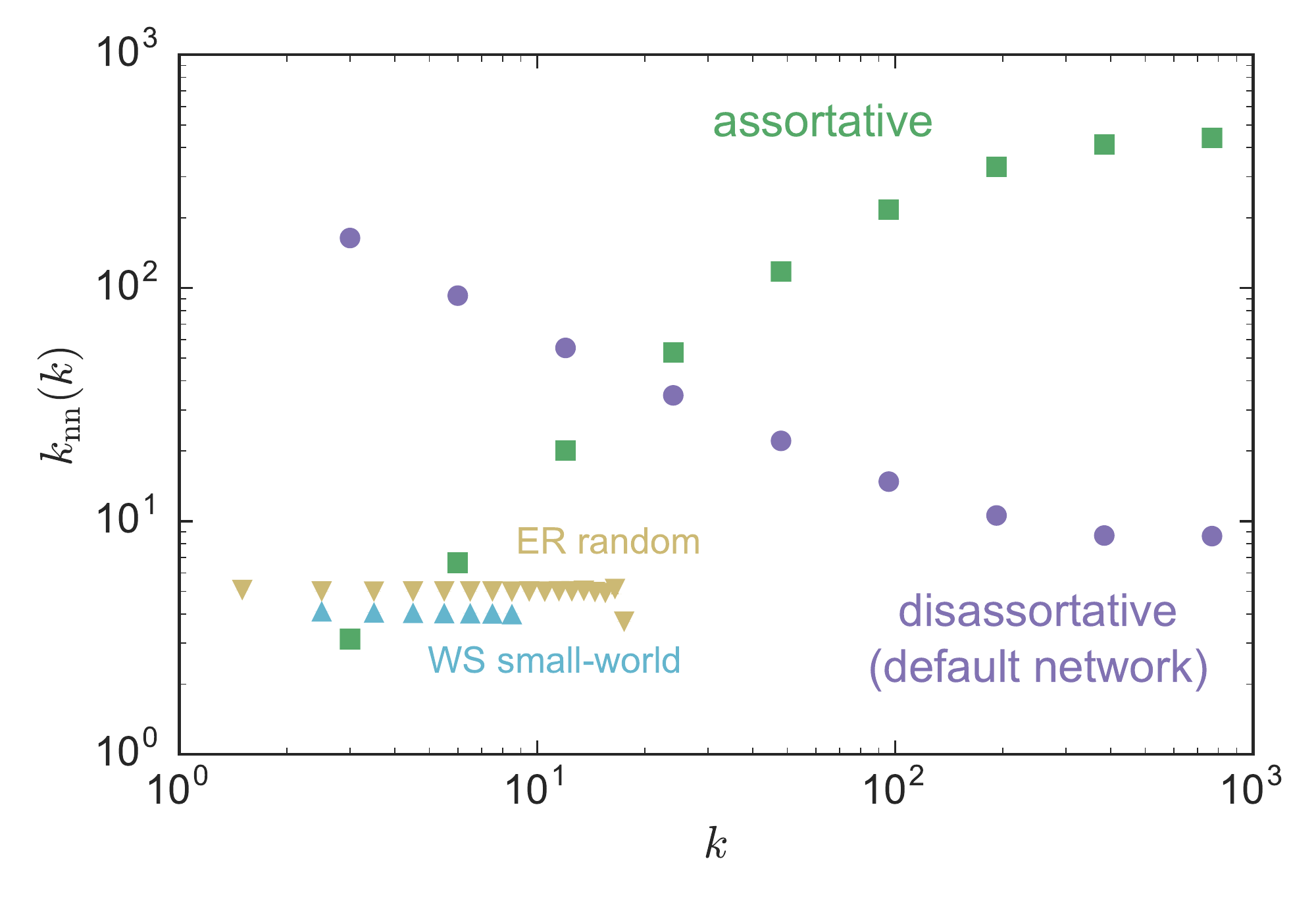}
    \caption{Average nn-degree $\langle k_{\mathrm{nn}}(k)\rangle$ vs degree $k$ for a disassortative network (default network) (purple circles), an assortative network created by reshuffling the links (green squares) \cite{Brunet:2004}, a  Erd\H os-R\`enyi (ER) random network (yellow triangles), and a Watts-Strogatz (WS) small-world network (blue triangles). Note that $\langle k_{\mathrm{nn}}(k)\rangle$ is grouped into bins of powers of $2$ and averaged within the bins for the scale-free networks. A bin for each degree is used for the ER random and WS small-world networks.}
    \label{knn}
  \end{minipage}
\end{figure}

Our network model has predominantly disassortative correlations (due to the scale-free exponent $\alpha<3$ \cite{Barrat:2005}) --- meaning that low-$k$ nodes tend to connect to high-$k$ nodes, and that the average nn-degree decreases with degree \cite{Newman:2002}. We see this in Figure~\ref{knn}, where we plot the average nn-degree $\langle k_{\mathrm{nn}}(k)\rangle$ as a function of degree for our network. The purple points indicate our preferential attachment model network, and we see that the average nn-degree is inversely related to the degree.  

The green curve shows a rewired assortative network \cite{Newman:2002} made by preserving the degrees of the original network but swapping links. To do this we use the method of Brunet \textit{et al}, using $N^2$ rewiring iterations with a parameter $p=0.99$ \cite{Brunet:2004}.  By modifying the nn-degrees of low degree nodes, we can investigate whether $k_{\mathrm{nn}}$ causes or is just correlated with informative low-$k$ nodes. Note that we use only the largest connected component of the rewired network, with $\langle N \rangle=9989$ nodes over 10 network realizations. 

The yellow triangles in Figure \ref{knn} show an Erd\H os-R\`enyi random network (ER). A random network is created by starting with $N$ nodes, and randomly connecting each pair of nodes with probability $p_{\mathrm{attach}} = \langle k \rangle/(N - 1)$ \cite{Barabasi:2016}. This results in a (peaked) binomial degree distribution, and completely uncorrelated connections where $k_\mathrm{nn} = \langle k^2\rangle/\langle k \rangle$ which is independent of individual node degree.  As before, we only use the  largest connected component, with $\langle N \rangle =9805$ nodes over 10 network realizations. The ER network also allows us to explore whether the heavy tail of the scale-free degree distribution is required to recover our observational results.

The light blue triangles in Figure \ref{knn} show a Watts-Strogatz (WS) small-world network \cite{Watts:1998}. This network starts with a uniform ring network with $k_i = \langle k \rangle$ for all nodes, and randomly rewires each link with probability $p_{\mathrm{rewire}}$ to another randomly selected node. We use $p_{\mathrm{rewire}} = 0.05$ to get the effects of both high clustering (i.e. links between neighbors of nodes) and short average path-lengths between arbitrary pairs of nodes \cite{Barabasi:2016}. This network has a narrowly peaked degree distribution, with a rapidly decaying exponential tail. ER and WS networks are similar, as both have short average path lengths between arbitrary nodes and non-heavy-tailed degree distributions, but the WS small-world network also has high clustering for small $p_{\mathrm{rewire}}$.

To examine network effects on our network aging model, we have kept the same model parameters for the (default) preferential attachment disassortative network, the assortative network, the ER random network, and the WS small-world network. (The scale-free exponent $\alpha$ is only used in the disassortative and assortative networks.) We examine $10$ random realizations of each network. We have also varied model parameters independently for each of these networks (data not shown) and obtain the same qualitative results.

\begin{figure} 
  \begin{minipage}[htb]{0.45\textwidth}
    \includegraphics[trim=15mm 13mm 5mm 5mm, clip,width=\textwidth]{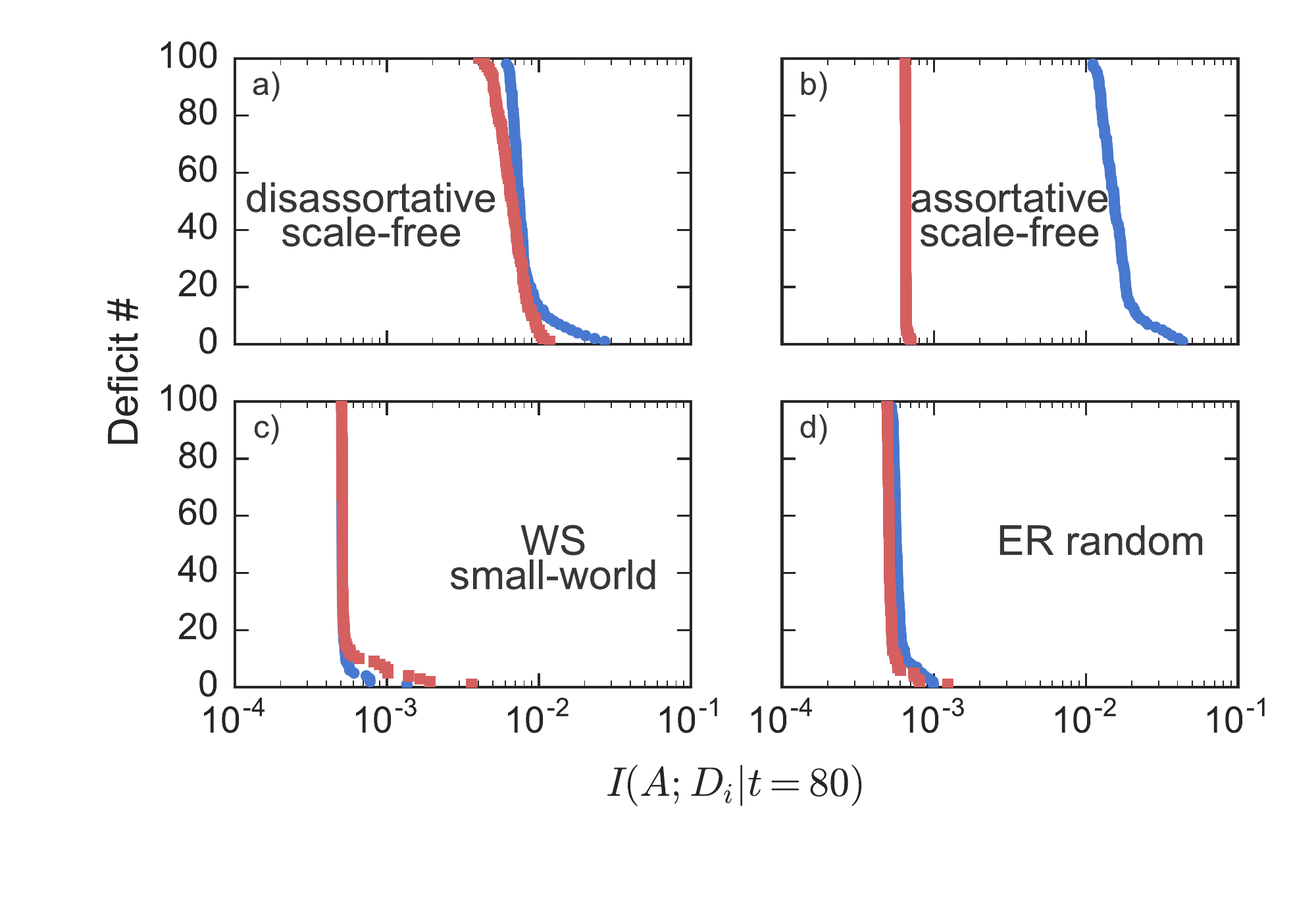}
    \caption{Rank ordered information $I(A;D_i|t=80)$ for the different networks, as indicated. The top $100$ most connected nodes in the network are in blue circles, and $100$ randomly selected nodes of the lowest degrees are in red squares. Results for each different network topology are averaged over $10$ randomly generated network realizations.}
    \label{Networks Fingerprint}
  \end{minipage}
\end{figure}

In Fig.~\ref{Networks Fingerprint} we show rank ordered information fingerprints for individual deficits $I(A;D_i|t)$, for the different network topologies as indicated. We observe striking differences in the scale and range of the mutual information with respect to mortality, and in the differences between the most and least connected nodes.  The random and small-world network both have a significantly smaller scale of mutual information, together with a much smaller range of variation. 

The scale-free disassortative (default) and assortative networks both have significantly higher scale of information for the most connected nodes, as well as considerable variation (approximately 10-fold) among them. However, while the disassortative network exhibits similar scales of information between the most and least connected nodes the assortative network does not. Furthermore, the assortative network shows only minimal variation of information among its least connected nodes.

Only the disassortative (default) network exhibits the fingerprint of mutual information of the NHANES and CSHA observational studies, in Figs.~\ref{NHANES Information Fingerprint} and \ref{CSHA Information Fingerprint} respectively: with considerable variation of mutual information between deficits, overlapping ranges between lab (low) and clinical (high) connectivity deficits, and mutual information on the order of $10^{-2}$ for individual deficits.  

\begin{figure} 
  \begin{minipage}[thb]{0.45\textwidth}
    \includegraphics[trim=10mm 10mm 3mm 5mm, clip,width=\textwidth]{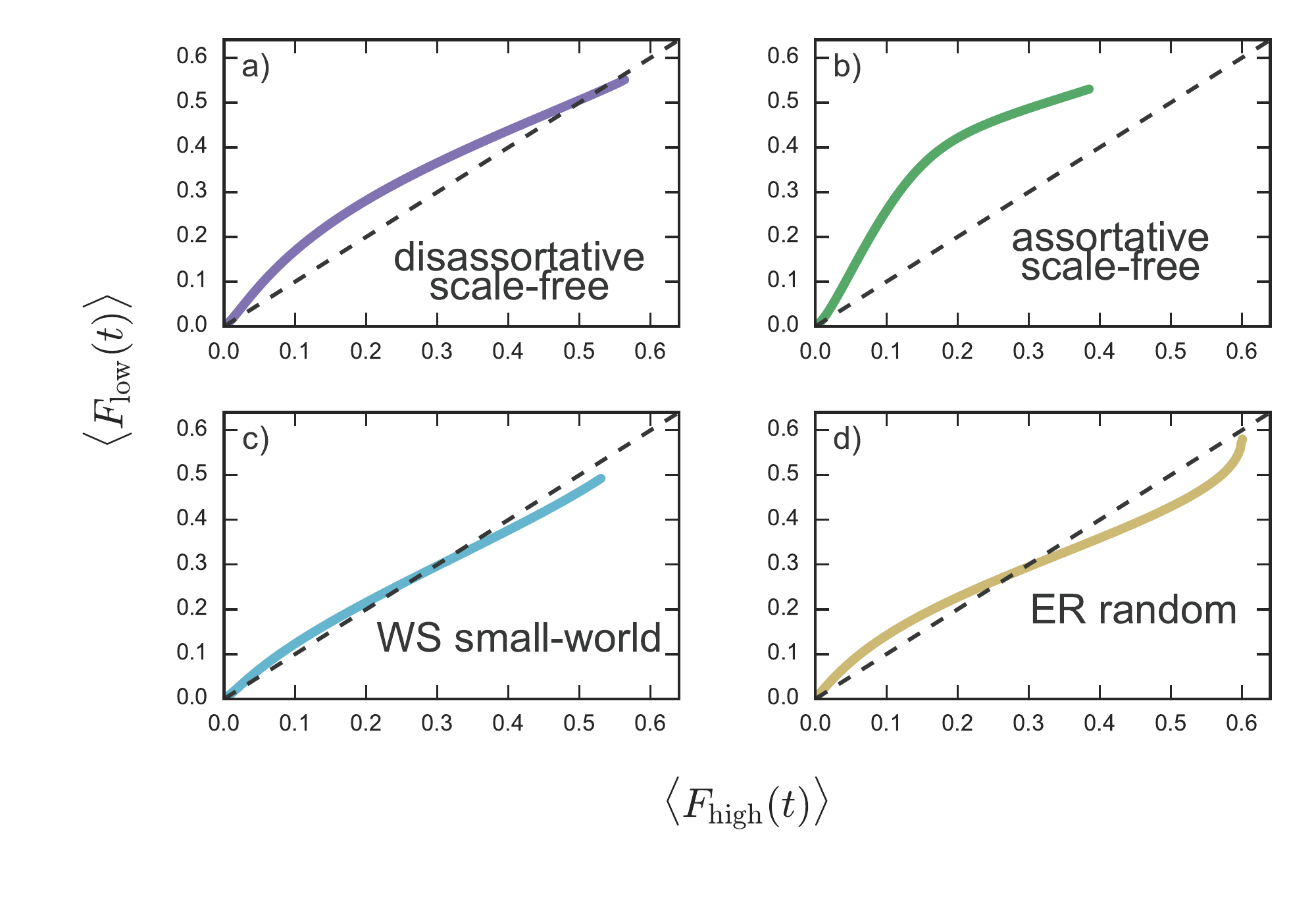}
    \caption{Average low-$k$  $\langle F_{\mathrm{low}}(t)\rangle$ vs average high-$k$  $\langle F_{\mathrm{high}}(t)\rangle$ plotted for $t = 0$ to $t = 110$ for our default network parameters (purple), the shuffled assortative network (green), the Erd\H os-R\`enyi random network (yellow), and the Watts-Strogatz small world network (light blue). The dashed black line shows the line $\langle F_{\mathrm{low}}(t)\rangle = \langle F_{\mathrm{high}}(t)\rangle$. Results are averaged over $10$ randomly generated networks and the standard deviations are smaller than the line width.}
    \label{Shuffled F vs F}
  \end{minipage}
\end{figure}

In Fig.~\ref{Shuffled F vs F}, we investigate the age-structure of the FIs generated by the low and high connectivity nodes. We plot $\langle F_{\mathrm{low}}(t) \rangle$ vs $\langle F_{\mathrm{high}}(t) \rangle$ for the different network topologies.  We see that the assortative network shows a rapid increase in $F_{\mathrm{low}}$, followed by growth of $F_{\mathrm{high}}$. In contrast, for the disassortative, random, and small-world networks there is comparable growth of both $F_{\mathrm{low}}$ and $F_{\mathrm{high}}$, though with higher $F_{\mathrm{low}}$ and a later cross-over for the disassortative network. 

\begin{figure} 
  \begin{minipage}[htb]{0.45\textwidth} 
    \includegraphics[trim=10mm 10mm 6mm 6mm, clip,width=\textwidth]{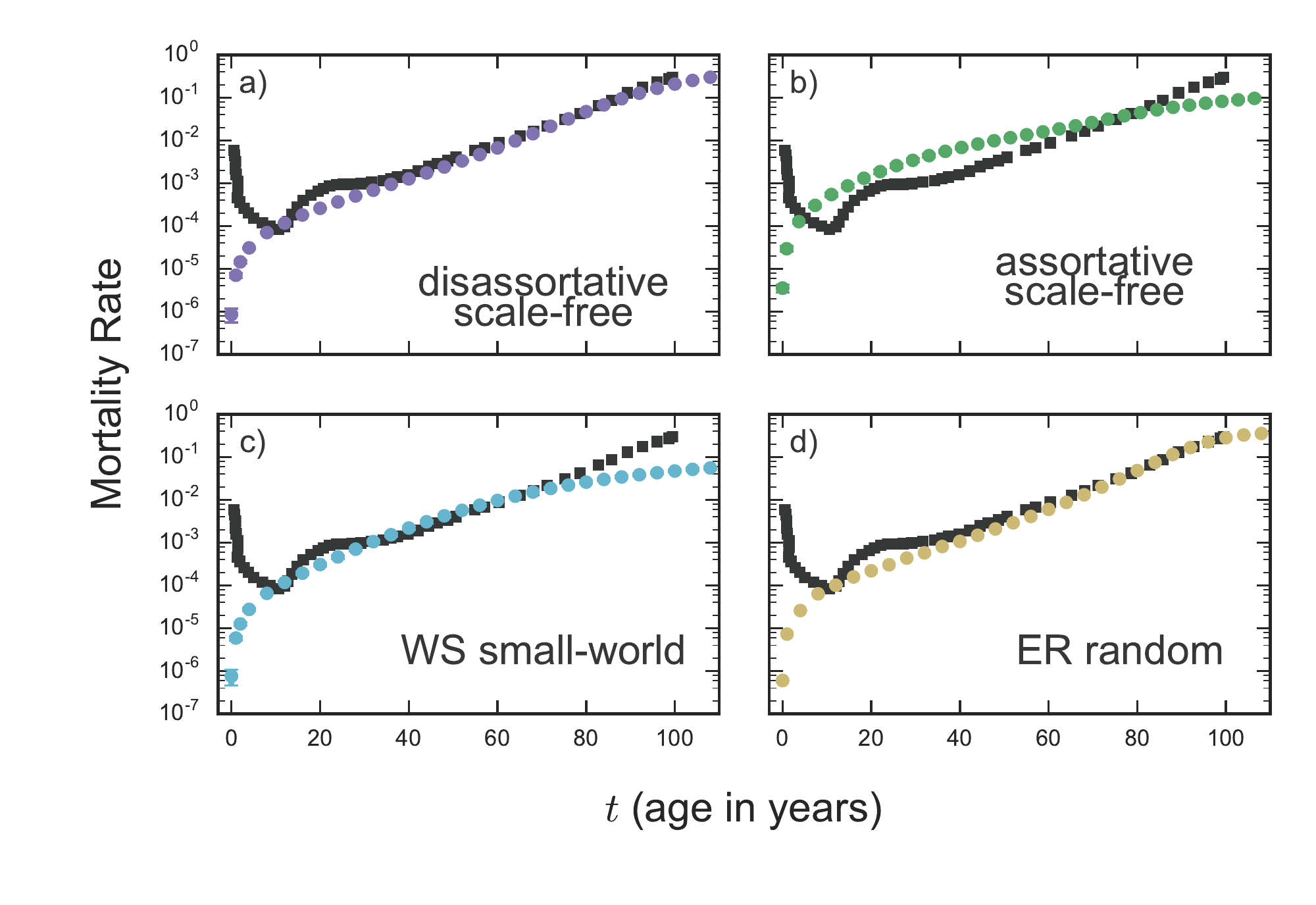}
    \caption{Mortality rate vs age for each of the networks. a) Disassortative scale-free network (purple circles), b) assortative scale-free network (green), c) WS small-world network (light blue), and d) ER random network (yellow). Computational results (circles) are averaged over $10$ randomly generated networks and error bars show the standard deviations. Black squares are observed human mortality rates \cite{Arias:2014}.}
    \label{Mortality Rates}
  \end{minipage}
\end{figure}

\begin{figure} 
  \begin{minipage}[htb]{0.45\textwidth} 
    \includegraphics[trim=10mm 10mm 6mm 6mm, clip,width=\textwidth]{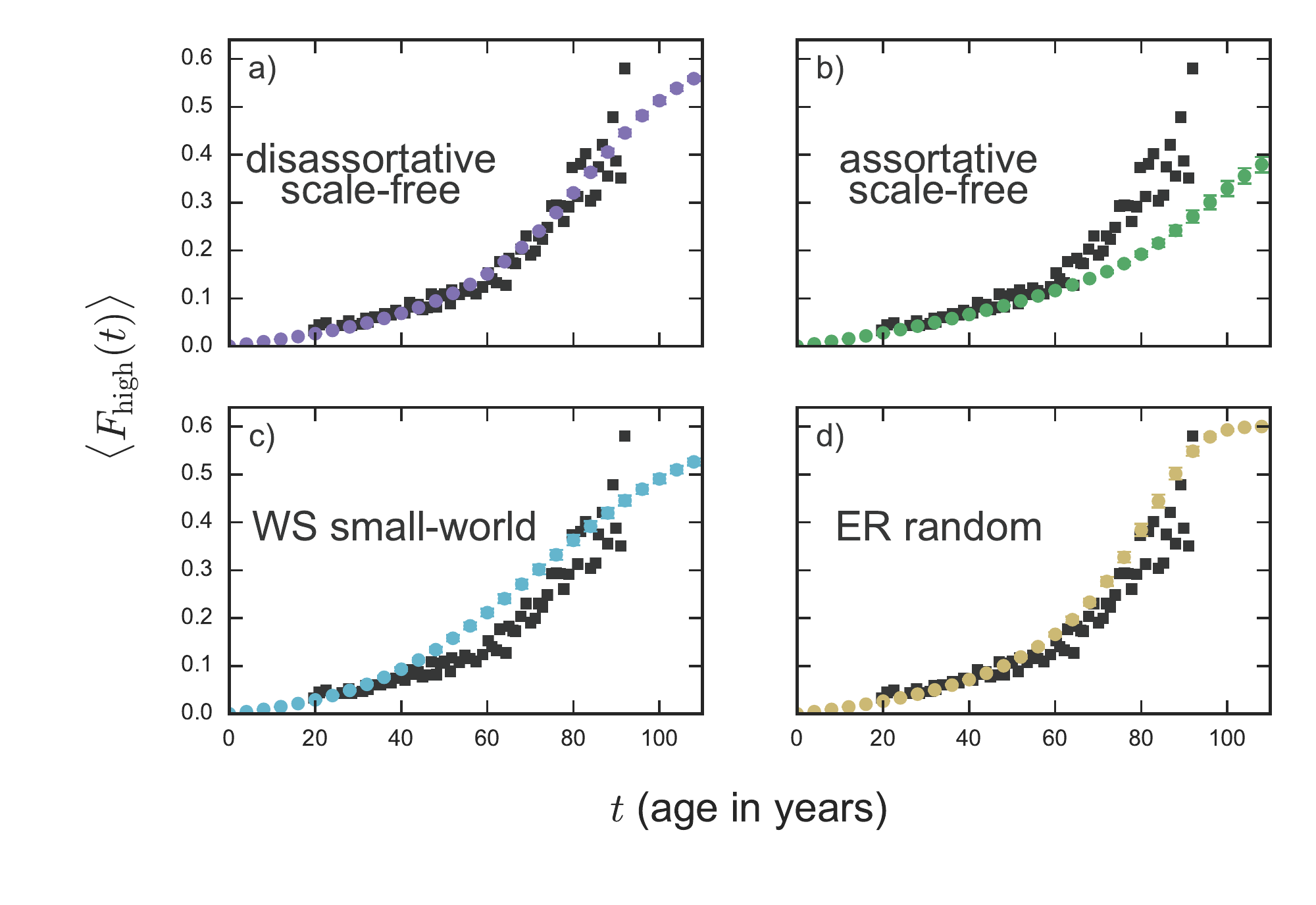}
    \caption{$\langle F_{\mathrm{high}} \rangle$ vs age for each of the networks, as indicated. a) Disassortative scale-free network (purple circles), b) assortative scale-free network (green), c) WS small-world network (light blue), and d) ER random network (yellow). Computational results (circles) are averaged over $10$ randomly generated networks and error bars show the standard deviations. Black squares are observed human clinical frailty \cite{Mitnitski:2013}.}
    \label{Frailty Growth}
  \end{minipage}
\end{figure}

In Fig.~\ref{Mortality Rates} we plot the average mortality rates vs age for different network topologies, with colored circles showing the computational model results and colored lines for the corresponding mean-field model results. Black squares indicate observed mortality rates \cite{Arias:2014}. Similarly, in Fig.~\ref{Frailty Growth} we plot $\langle F_{\mathrm{high}} \rangle$ vs age $t$ for both observational data (black squares) and model data for different networks (coloured points). 

Even without parameter adjustment, most of the network topologies approximately capture the observational data after $t=20$ years. Some differences are seen, particularly for the assortative scale-free network in the mortality rate. This agreement indicates that mortality and frailty data alone do not strongly constrain the network topology.

From Fig.~\ref{Networks Fingerprint}, we observed early damage of $F_{\mathrm{low}}$ in the assortative network. Our MFT allows us to narrow down what aspects of the network are leading to this behavior, since the only aspects of the network structure included are the degree distribution $P(k)$ and nn-degree correlations $P(k'|k)$. 

Different network topologies are easily introduced provided $P(k)$ and $P(k'|k)$ are known. The exact $P(k)$ for our default shifted-linear preferential attachment networks \cite{Fotouhi:2013}, ER random networks, and WS small-world networks \cite{Barrat:2000} are known. (We remove zero degree nodes from the ER random degree distribution, so that $P_{k\neq 0}(k) = P(k)/\sum_{l \neq 0}P(l)$.) Using various $P(k'|k)$ we can then put different degree correlations into our MFT network. We include three types of degree correlations, uncorrelated (neutral), assortative, and disassortative \cite{Barabasi:2016}.

For a network with uncorrelated (neutral) connections, $P(k'|k) = k'P(k')/\langle k \rangle$. We then have $k_{\mathrm{nn}}(k) = \sum_{k'} k' P(k'|k) = \langle k^2 \rangle/\langle k \rangle$, so that all nodes have the same nn-degree.  These correlations are used for ER random and WS small-world networks, and recover the approximately constant $k_{\mathrm{nn}}$ that we observed in Fig.~\ref{knn}.

In a network with assortative correlations, nodes tend to be connected to other nodes of similar degree. Assortative correlations that approximate those used in our computational model in Sec.~\ref{network structure} are \cite{Moreno:2003} $P(k'|k) = \alpha \delta_{k'k} + (1 - \alpha) k'P(k') /\langle k \rangle$. These lead to, $k_{\mathrm{nn}}(k) = \sum_{k'} k' P(k'|k) = \alpha k + (1 - \alpha) \langle k^2 \rangle /\langle k \rangle$, which increases linearly with $k$ (see Fig.~\ref{knn}). Changing $\alpha$ modifies the amount of assortative correlation; we use $\alpha = 0.8$.

In a network with disassortative connections, nodes tend to be connected to other nodes of differing degree. The (disassortative) correlations for our default shifted-linear preferential attachment network are \cite{Fotouhi:2013},
\begin{gather} \nonumber
P(k'|k) = \frac{ \Gamma(k + \lambda + \alpha) \Gamma(k' + \lambda) }{ k \Gamma(m + \lambda) \Gamma(k + k' + 2\lambda+\alpha) } \\\times \Bigg[ \sum_{i=m+1}^k \frac{\Gamma(i + m + 2\lambda + \alpha - 1)}{\Gamma(i + \lambda + \alpha - 1)}\binom{k+k'-m-i}{k'-m} \\ \nonumber + \sum_{i = m+1}^{k'}\frac{\Gamma(i + m + 2\lambda + \alpha - 1)}{\Gamma(i + \lambda + \alpha - 1)}\binom{k+k'-m-i}{k-m} \Bigg],  \label{default correlations}
\end{gather} 
where $m = \langle k \rangle/2  = k_{\mathrm{min}}$ and $\lambda  = m (\alpha - 3)$. This is exact in the limit $N \to \infty$ \cite{Fotouhi:2013}, and gives disassortative correlations where $k_{\mathrm{nn}}(k)$ decreases with $k$.

\begin{figure} 
  \begin{minipage}[thb]{0.45\textwidth}
    \includegraphics[trim=3mm 5mm 4mm 3mm, clip,width=\textwidth]{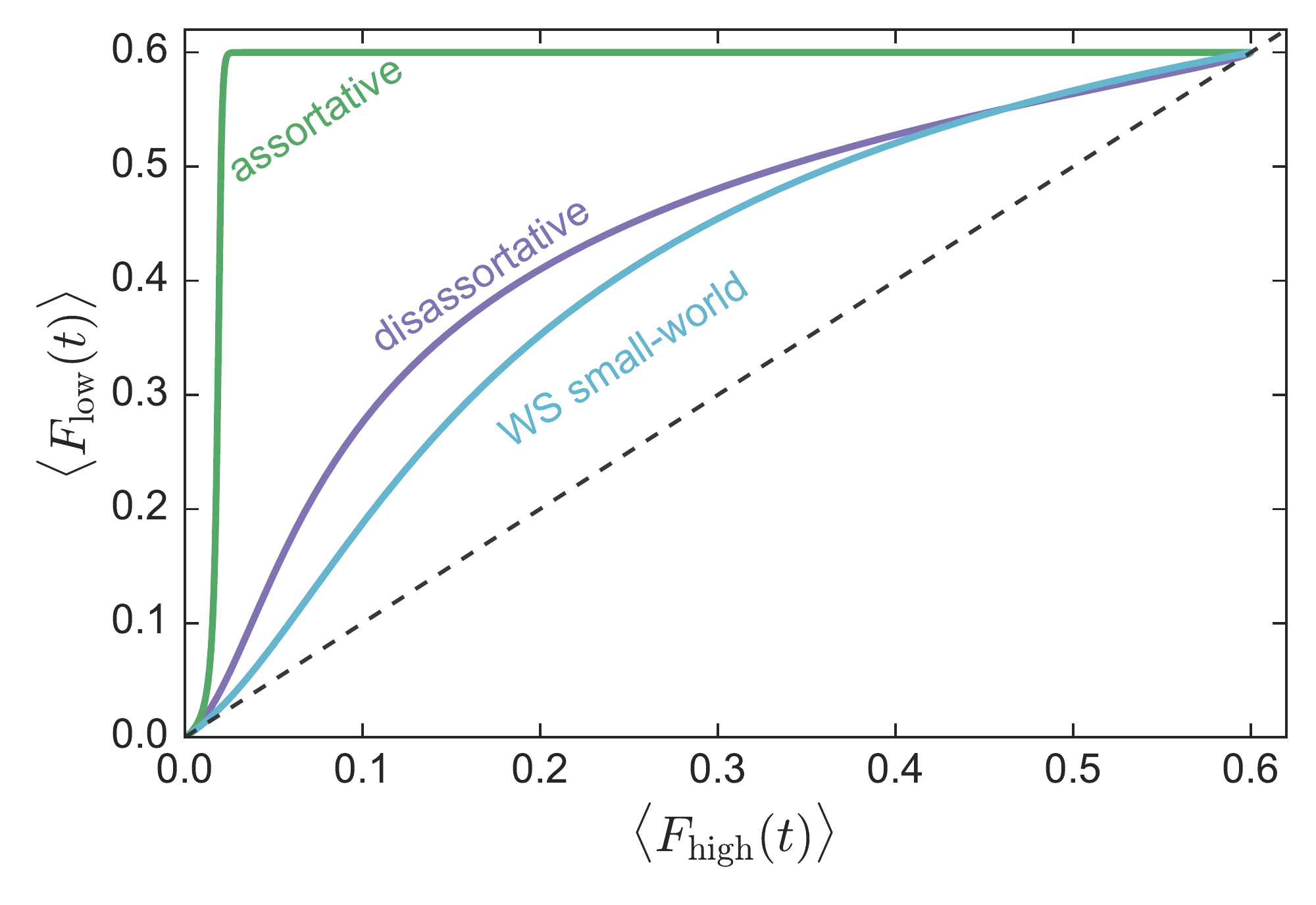}
    \caption{Average low-$k$  $\langle F_{\mathrm{low}}(t)\rangle$ vs average high-$k$  $\langle F_{\mathrm{high}}(t)\rangle$ from our mean-field model in Sec.~\ref{MFT}. The dashed black line shows the line $\langle F_{\mathrm{low}}(t)\rangle = \langle F_{\mathrm{high}}(t) \rangle$. A scale-free network with preferential attachment disassortative correlations (default network) in purple, scale-free network with assortative correlations in green, and a WS small-world network with neutral correlations in light blue.}
    \label{Mean Field F vs F}
  \end{minipage}
\end{figure}

In Fig.~\ref{Mean Field F vs F} we show the average low-$k$ FI vs the average high-$k$ FI, $\langle F_{\mathrm{low}}(t)\rangle$ vs $\langle F_{\mathrm{high}}(t)\rangle$ from our MFT. In purple we use the (default) preferential attachment disassortative correlations, in green we use assortative correlations, and in light blue we use a WS small-world network. We see qualitative agreement with the age-structure shown in Fig.~\ref{Shuffled F vs F} -- confirming that nn-degree correlations (included in our MFT) are important for the observed age-structure. [We have not shown MFT results for the ER random network since $\langle F_{\mathrm{low}} \rangle$ behaves poorly when it includes nodes with $k \leq 2$, due to their great variability of local frailty $f_i$.]

\subsection{Mutual information of FI with mortality} 
\begin{figure} 
  \begin{minipage}[thb]{0.45\textwidth}
    \includegraphics[trim=3mm 5mm 4mm 3mm, clip,width=\textwidth]{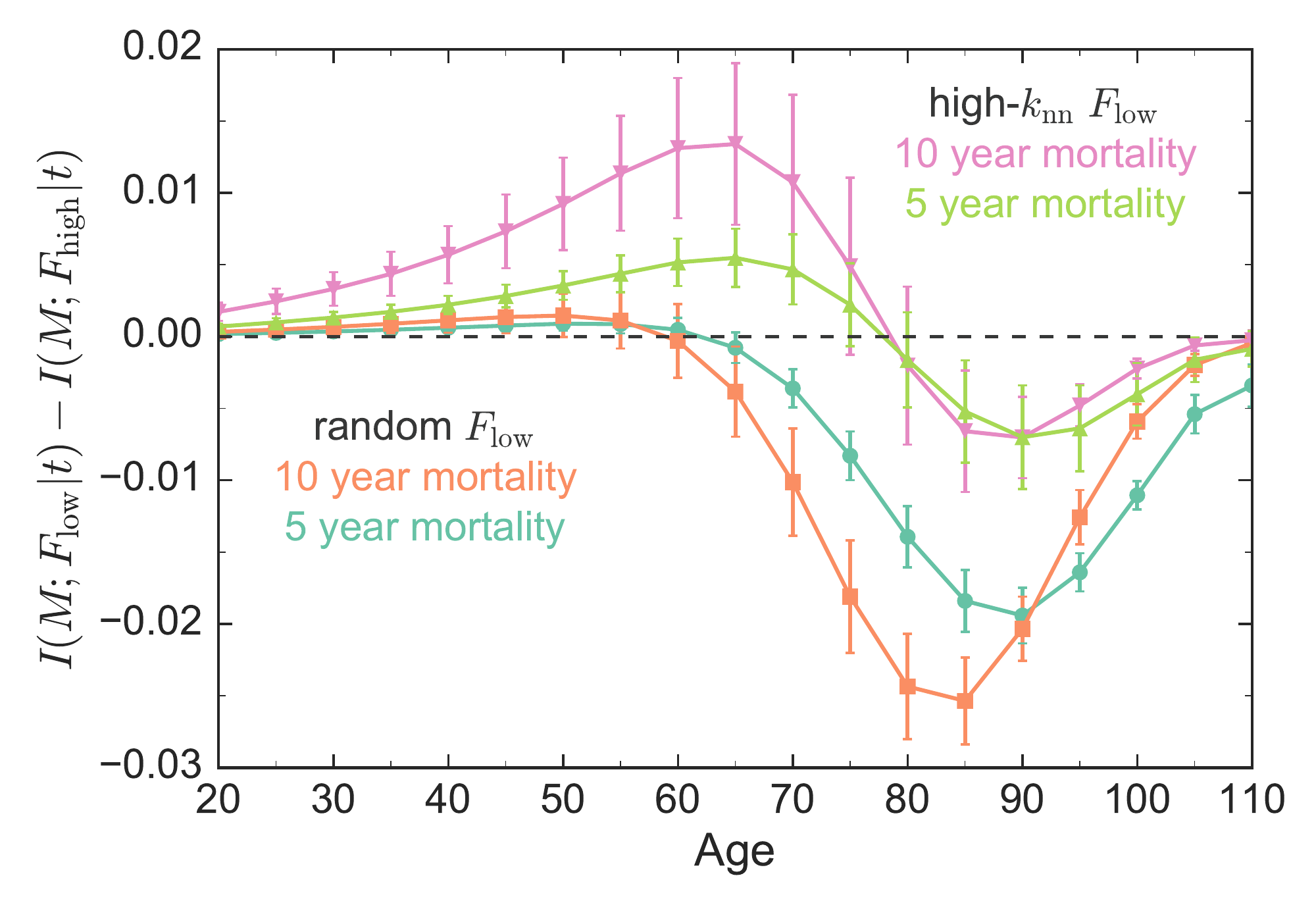}
    \caption{The difference in mutual information of $F_{\mathrm{low}}$ and $F_{\mathrm{high}}$ ( $I(M;F_{\mathrm{low}}|t)- I(M;F_{\mathrm{high}}|t)$ ) vs age $t$ for different binary mortality outcomes. 5 year mortality outcomes are shown as turquoise circles and 10 year as orange squares. The dashed line shows when the information of both FIs are equal. Error bars represent the standard deviation between randomly generated networks.  The purple down and green up triangles indicate the information difference for 10 and 5 year mortality, respectively, of $F_{\mathrm{low}}^{\mathrm{high}\text{-}k_{\mathrm{nn}}}$ which is constructed with $n=32$ nodes with $k=2$ that are randomly chosen from those with above-average $k_{\mathrm{nn}}$.}
    \label{Finformation}
  \end{minipage}
\end{figure}

We have seen that $F_{\mathrm{low}}$ damages earlier than $F_{\mathrm{high}}$ (Fig.~\ref{Model Time Structure}) and that the mutual information of poorly connected ($k=2$) nodes with large nearest-neighbor degree significantly overlaps with the informativeness of the most connected nodes (Fig.~\ref{Model Spectra}) in our (disassortative) scale free network model. Because of these informative earlier damaged nodes, we were interested in whether $F_{\mathrm{low}}$ could be more informative of mortality than $F_{\mathrm{high}}$, particularly at younger ages. In Fig.~\ref{Finformation} we show the difference in information for $F_{\mathrm{low}}$ and $F_{\mathrm{high}}$ for different mortality outcomes vs age. We find that $F_{\mathrm{low}}$ is slightly more informative at ages less than $\approx 65$ and is increasingly more informative than $F_{\mathrm{high}}$ at these younger ages for longer mortality outcomes. This is the result of $F_{\mathrm{low}}$ nodes damaging early but having a delayed effect on mortality, so that they are an early predictor of later mortality, but not so much immediate mortality. The relatively large standard deviations for different randomly generated networks shows that this result is affected by the particular randomly generated network.

While the observational NHANES and CSHA sample-sizes are much smaller, a similar calculation shows a slightly lower $F_{\mathrm{lab}}$ information $-0.002 \pm 0.013$ compared to $F_{\mathrm{clin}}$ in the NHANES data for younger people ($65-75$ years) and a slightly higher mutual $F_{\mathrm{lab}}$ information $+0.033 \pm 0.027$ compared to $F_{\mathrm{clin}}$  in the CSHA data. While we do not have sufficient data to vary our mortality outcome to determine if $F_{\mathrm{lab}}$ is more predictive of later mortality outcomes as we did in the model, we can see  in the CSHA data that $F_{\mathrm{lab}}$ is more informative for younger people.  

Since we found that the most informative low-connectivity nodes were those with large $k_{\mathrm{nn}}$, we also considered an FI constructed from $n=32$ randomly chosen nodes of lowest degree ($k=2$) from those that have above-average  $k_{\mathrm{nn}}$. The information advantage of  $F_{\mathrm{low}}^{\mathrm{high}\text{-}k_{\mathrm{nn}}}$ is indicated in Fig.\ref{Finformation} with down and up triangles for 10 and 5 year mortality, as indicated. The advantage over $F_{\mathrm{high}}$ is large and significant for ages below $t=80$ years, with a stronger advantage at earlier ages for later mortality. This will be an attractive avenue to pursue.

\section{Summary and Discussion} 

The observational $F_{\mathrm{clin}}$ or $F_{\mathrm{lab}}$ respectively measure clinically observable damage that tends to occur late in life or pre-clinical damage that is typically observable in lab tests or biomarkers before clinical damage is seen. However, they are similarly informative of human mortality \cite{Howlett:2014, Blodgett:2017, Blodgett:2016}. Our analysis indicates that individual laboratory and clinical deficits have broad and overlapping ranges of mutual information. 

Our working hypothesis is that clinical deficits correspond to high connectivity nodes of a complex network, while laboratory deficits correspond to lower connectivity nodes. With our network model of individual aging and mortality, we have confirmed that $F_{\mathrm{high}}$ and $F_{\mathrm{low}}$, formed from high and low connectivity nodes respectively, behave similarly to the observational $F_{\mathrm{clin}}$ and $F_{\mathrm{lab}}$.

Within the context of our aging model, we uncover the mechanisms of this observed behavior. In our model low-$k$ nodes tend to damage before high-$k$ nodes. This is because of the larger average damage rates of low-$k$ nodes compared to high-$k$ nodes (as calculated with our network mean field theory, and illustrated in Fig.~\ref{Damage Rates Figure}). At the same time, our information spectrum shows that information $I(A;D_i|t)$ increases with $k$. Roughly speaking, high-$k$ nodes need a larger local frailty $f$ to have comparable damage rates as low-$k$ nodes.  Thus, damage of high-$k$ nodes is informative of high network damage, which also leads to mortality. This is why high-$k$ nodes both damage later and are informative of mortality (Fig.~\ref{Model Spectra}b).

However, some low-$k$ nodes also damage later and are highly informative of mortality. Information $I(A;D_i|t)$ increases with $k_{\mathrm{nn}}$ for the low-$k$ nodes, and low-$k$ high-$k_{\mathrm{nn}}$ nodes damage later. This can also be explained using the network structure. Low-$k$ nodes are protected from damage when they are connected to high-$k$ nodes.  Rapidly damaging low-$k$ nodes without this protection tend to damage early for most individuals, giving these nodes a low information value of mortality. Conversely, protected nodes tend to damage only when their high degree neighbors start to damage, which only occurs when the network is  heavily damaged and close to mortality. As a result, only the low-$k$ nodes with high-$k_{\mathrm{nn}}$ are highly informative (Fig.~\ref{Model Spectra}a). Interestingly these nodes still tend to damage before high-$k$ nodes, leading to an early predictor of mortality.

Degree correlations control the average degree of neighboring nodes and hence control the amount of protection in low-$k$ nodes.  By modifying the degree correlations in the network in our computational model we have shown that this protection can be caused by disassortative correlations --- where low-$k$ nodes tend to attach to high-$k$ nodes. Conversely, eliminating low-$k$ high-$k_{\mathrm{nn}}$ nodes by modifying the network to introduce assortative correlations removes this protection, and we then find all low-$k$ nodes have low information (Fig.~\ref{Networks Fingerprint}b). 

Our mean-field model allows us to explicitly modify the degree distribution and the degree correlations with the nearest-neighbor degree distribution $P(k'|k)$, and to include no other network features. In our mean-field model we see similar results to our computational model where, e.g., adding assortative correlations increases the rate at which $F_{\mathrm{low}}$ increases with respect to $F_{\mathrm{high}}$. This confirms that degree distribution and degree correlations largely determine the early damage of low-$k$ nodes that we observe in  scale-free networks.

Degree distributions and correlations only weakly control the behavior of ER random and WS-small world networks. The low variation in $k$ and $k_{\mathrm{nn}}$ in those networks results in a lack of contrast between the damage rates of nodes. This leads to node information that is nearly constant throughout the network and to only small differences in the damage structure of low-$k$ and high-$k$ nodes (Fig.~\ref{Networks Fingerprint}c and d). This also leads to low magnitude of the mutual information per node, since nodes behave much more uniformly and ``randomly'' than in a scale-free network. However, we can still see some protection in low-$k$ nodes. This is particularly apparent in the ER random network when $F_{\mathrm{high}}$ surpasses $F_{\mathrm{low}}$ (Fig.~\ref{Shuffled F vs F}d). 

The behavior of observational deficits seems to best resemble the behavior of the computational model with a scale-free network and disassortative correlations. Node information seen in the (default) scale-free disassortative network is a much better qualitative match of observational data, as compared with scale-free assortative, WS small-world, or ER random networks. 

Our analogy between observational deficits and model nodes allows us to make predictions about the underlying network structure of observational health deficits, even though we cannot directly measure this network. The observational network should have a heavy-tail degree distribution, so that a large range of possible information values can be obtained. The network should also include disassortative correlations so that there are connections between high-$k$ and low-$k$ nodes, allowing low-$k$ nodes to be informative of mortality. 

From observational data we find that clinical deficits that integrate many systems into their performance (e.g. functional disabilities, or social engagement) are very informative (Figs.~\ref{NHANES Information Fingerprint} and \ref{CSHA Information Fingerprint}). In contrast, single diagnoses, even ones strongly associated with age such as osteoporosis, on their own offer less value. The model interpretation of this is that these high information disability deficits have a higher connectivity than lower information clinical deficits. It intuitively makes sense for deficits that integrate many systems to have a large connectivity. In support of this, our partial network reconstruction (Fig.~\ref{Reconstructed Clinical Fingerprint}) shows that high information clinical deficits in both the NHANES and CSHA correspond to nodes with a high reconstructed degree.   

We have shown that the age-structure of network damage is related to the network structure. Highly informative low-degree nodes (pre-clinical deficits) damaged early in life promote the damage of their high-degree neighbors, but the damage to their high-degree neighbors takes time and is not seen in the high-degree (clinical) FI until later ages. Indeed, we have shown that a $F_{\mathrm{low}}$ is slightly more informative at earlier ages, and is increasingly informative for longer mortality outcomes ($5$ year vs $10$ year) (see Fig.~\ref{Finformation}). Choosing more high-$k_{\mathrm{nn}}$ nodes in $F_{\mathrm{lab}}$ significantly enhances this  effect. Low-$k$ nodes are informative of long-term mortality rather than short-term. Similar results are seen in the observational CSHA data, which indicates that $F_{\mathrm{lab}}$ could  be used as an early measure of risk of future poor health.

Our network model is generic, without a specific mapping between model nodes and observed human deficits. This is because we have no reliable way of extracting a specific network from observational data, though we have shown that rank-ordering of high-connectivity nodes can be done. This is also because distinct parameterization of every node of such a network model would require enormous amounts of observational data, if it could be done at all.  Nevertheless, we can used our generic model to explore robust qualitative phenotypes --- to uncover generic mechanisms, to predict behavior, and to improve the utility of the Frailty Index in human aging and mortality. 

In this paper we have kept our model parameterization unchanged from the default parameters, though we have checked (data not shown) that our results are qualitatively robust to parameter variation.  This has allowed us to explore the impact of network topology on mortality statistics (a small effect) and on mutual information between health deficits (a strong and distinctive effect). The  $F_{\mathrm{high}}$ and $F_{\mathrm{low}}$ model phenomenology are also affected by  changes in network topology. This indicates that both $F_{\mathrm{high}}$ and $F_{\mathrm{low}}$ are usefully distinct characteristics of health in our network model. Our results provide insight into the mechanisms of the similarly useful and distinct observational $F_{\mathrm{clin}}$ and $F_{\mathrm{lab}}$ \cite{Blodgett:2016, Howlett:2014, Mitnitski:2015}. 

\subsection{Acknowledgments} 
We thank ACENET and Compute Canada for computational resources. ADR thanks the Natural Sciences and Engineering Research Council (NSERC) for operating grant RGPIN-2014-06245.  KR is funded in this work by career support as the Kathryn Allen Weldon Professor of Alzheimer Research from the Dalhousie Medical Research Foundation, and with operating funds from the Canadian Institutes of Health Research (MOP-102544) and the Fountain Innovation Fund of the Queen Elizabeth II Health Science Foundation. SGF thanks NSERC for a CGSM fellowship. 

\bibliography{ref}
\end{document}